\documentclass[
reprint,
superscriptaddress,
amsmath,amssymb, 
aps,
pre,
]{revtex4-2}

\usepackage{microtype}
\usepackage{xurl}
\usepackage{etoolbox}

\AtBeginEnvironment{thebibliography}{%
  \raggedright
  \hyphenpenalty=500
  \exhyphenpenalty=500
}
\usepackage{ulem}
\usepackage[english]{babel}
\usepackage{dsfont}
\usepackage{amsmath}
\usepackage{hyperref}
\usepackage{mathrsfs}
\usepackage{bm}
\usepackage{cancel}
\usepackage{graphicx}
\usepackage{dcolumn}
\usepackage{amsmath}
\usepackage{placeins}
\usepackage{comment}
\usepackage{xcolor}
\usepackage{physics}
\usepackage{listings}
\usepackage{mhchem}

\usepackage{bbold}

\usepackage{booktabs}
\usepackage{lipsum}

\newcommand{\RN}[1]{%
  \textup{\uppercase\expandafter{\romannumeral#1}}%
}

\usepackage{lipsum}

\definecolor{sclr}{rgb}{0.27, 0.51, 0.71} 
\definecolor{BrighterTeal}{HTML}{005F56}
\definecolor{RichPlum}{HTML}{5B2A3F}
\definecolor{DeepBlue}{HTML}{2F4A72}
\definecolor{WarmOchre}{HTML}{A36D3E}

\begin{document}

\title{How gravity turns active, chemically-fueled droplets into oscillators }

\author{Advait Thatte}
\affiliation{Faculty of Mathematics, Natural Sciences, and Engineering: Institute of Physics, University of Augsburg, Universit\"atsstra\ss e~1, 86159 Augsburg, Germany}

\author{Judit Sastre}
\affiliation{Department of Bioscience, Technical University of Munich, Lichtenbergstra{\ss}e 4, 85748 Garching, Germany}

\author{Job Boekhoven}
\affiliation{Department of Bioscience, Technical University of Munich, Lichtenbergstra{\ss}e 4, 85748 Garching, Germany}

\author{Christoph A. Weber}
\email{christoph.weber@physik.uni-augsburg.de}
\affiliation{Faculty of Mathematics, Natural Sciences, and Engineering: Institute of Physics, University of Augsburg, Universit\"atsstra\ss e~1, 86159 Augsburg, Germany}

\date{\today}
\begin{abstract}
Oscillations are a fundamental form of temporal organization in living and synthetic matter, commonly attributed to nonlinear chemical reaction networks. 
However, whether oscillatory dynamics can emerge collectively from phase separation and transport under stationary chemical driving remains poorly understood. 
Here, we show how chemically active droplets can form a self-sustained oscillator through the interplay of gravity, fusion, and reaction-controlled droplet size selection. 
We develop a discrete-droplet framework coupling chemical reactions, diffusion, stochastic nucleation, and hydrodynamic transport, which quantitatively captures oscillations observed in chemically fueled coacervates confined within synthetic cells. 
Sedimentation of many droplets and gravity-induced fusion drive the underlying feedback mechanism, displacing droplets beyond their chemically selected stationary size. 
Their subsequent shrinkage releases material promoting nucleation of a new generation of sedimenting droplets at the top of the synthetic cell, closing the oscillatory cycle. 
We demonstrate that the oscillation period is governed by the combined timescales of nucleation and sedimentation, while nucleation kinetics control the emergence of coherent, propagating droplet pulses. 
These results identify a physical mechanism by which stationary chemical driving generates collective oscillations and spatiotemporal organization in active emulsions, without requiring oscillatory reaction kinetics. 
\end{abstract}

\maketitle

\section{Introduction}

\begin{figure*}[tb]  \includegraphics[width=\linewidth]{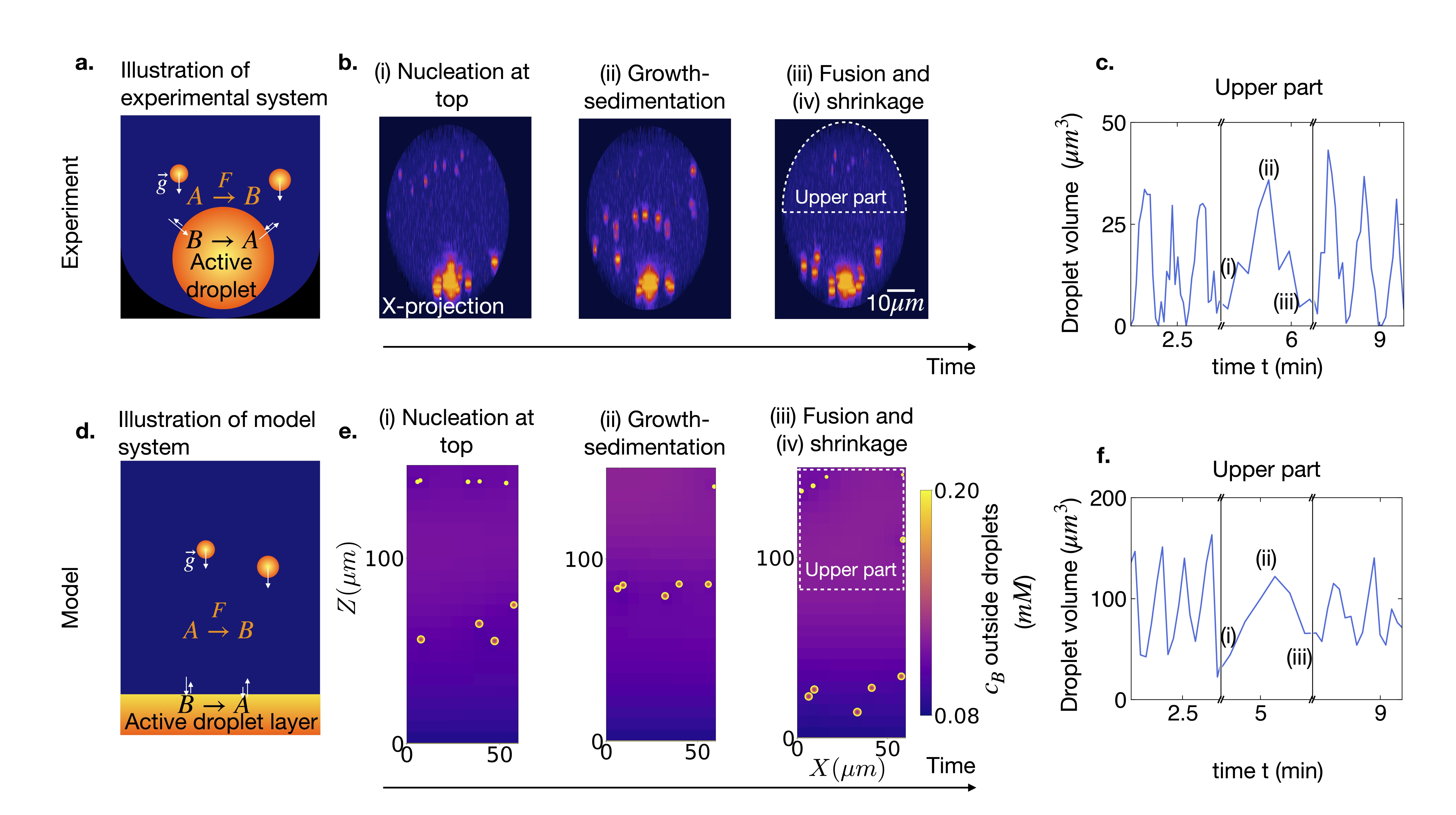}
    \caption{
    \textbf{
Chemically fueled droplets sediment under gravity, leading to oscillating droplet pulses.}
    (a,d) Sketch of the experimental (top, a) and model (bottom, d) system, with experimental details discussed in Ref.~\cite{sastre2025size}.
    Gravity confines the active droplet (simplified to a layer in the model) at the bottom where it wets. 
    The main component $B$ of the active droplet (orange) chemically deactivated inside to a less phase-separating state $A$, leading to efflux of $A$ into the outside phase (blue), where activation from $A$ to $B$ occurs through the hydrolysis of fuel $F$. 
    Both, experiments and model leads to oscillations of the active droplet (layer) size.
    Moreover, in both (b,e), droplets sediment collectively as a pulse. The cycle is composed of four different stages:
    (i) nucleation of small droplets at the top, (ii) their growth and sedimentation, and (iii) fusion at bottom with the bigger bottom droplets (layer), (iv) followed by its shrinkage. 
    The oscillation frequency and amplitude of the droplet volume in the top domain of the spherical {micro-reactor} show good quantitative agreement between the experiments and the rectangular box used in the model (f). For the experiments, the diameter of the spherical {micro-reactor} is around $50 \mu m$. Concentration of fuel $c_F$ is $3.3$mM and conserved quantity $\psi$ is $10$mM. We use the same concentrations $c_F$ and $\psi$ in the simulations as in the experiments. We use a box of height $H=150\mu m$ as shown in (e). 
    We chose a slightly larger {micro-reactor} height $H$ in the simulations than in the experiments to compensate for differences in the geometry of the rectangular {micro-reactor} in the model compared to the spherical {micro-reactor} in the experiments.}
    \label{fig:fig1}
\end{figure*}

Oscillatory dynamics are a hallmark of living matter, coordinating processes across a broad range of temporal and spatial scales.
Examples include circadian clocks~\cite{Hardin1990,NovakTyson2008}, calcium oscillations~\cite{Berridge1998}, glycolytic oscillations~\cite{Goldbeter1996}.
In systems chemistry, sustained oscillations are typically generated by nonlinear reaction networks containing feedback loops and separated kinetic timescales, with paradigmatic examples including the Belousov--Zhabotinsky reaction and other reaction--diffusion systems~\cite{EpsteinPojman1998,FieldBurger1985}.
Oscillatory reaction networks rely on the interplay of fast autocatalytic activation and slower inhibitory or depletion processes~\cite{NovakTyson2008, EpsteinPojman1998}.
The resulting delayed negative feedback prevents relaxation to equilibrium and instead drives  oscillations around an unstable steady state.

A further, equally fundamental organizing principle of living matter is liquid--liquid phase separation~\cite{Banani2017,ShinBrangwynne2017,AlbertiHyman2021}.
Biomolecular condensates and coacervate droplets compartmentalize biochemical reactions without membranes~\cite{Banani2017,Lyon2021} and are largely governed by thermodynamics near phase equilibrium~\cite{fritsch2021local}.
Consequently, droplets are remarkably robust objects: they nucleate, coarsen, fuse and eventually minimize their interfacial energy through Ostwald ripening~\cite{Weber2019,Zwicker2022}.
While this stability is advantageous for compartmentalization in cell biology~\cite{Banani2017,AlbertiHyman2021}, it also makes it difficult to realize controlled cycles of droplet growth and dissolution.
Generating sustained oscillations of droplets therefore requires mechanisms that continuously drive them away from thermodynamic 
equilibrium.

A growing body of work has therefore explored the interplay between chemical reactions and phase separation.
In one direction, researchers have used phase-separated compartments to regulate chemical reaction networks by selectively partitioning dilute chemical clients, thereby modifying reaction kinetics, reaction pathways, and even oscillatory reaction networks~\cite{Nakashima2019,Laha2024,Smokers2025}.
Conversely, chemical reactions have been exploited to regulate phase separation, enabling reaction-controlled droplet nucleation, suppression of Ostwald ripening and droplet size selection~\cite{Zwicker2015, Zwicker2017, Kirschbaum2021, Bauermann2025}, spontaneous division~\cite{Zwicker2017, Bauermann2022}, and vacuole formation~\cite{Donau2020, 
bartolucci2021controlling,Bergmann2023,Bauermann2023}.
More recently, oscillatory reaction networks have been coupled to phase separation to periodically assemble and dissolve droplets, thereby translating chemical oscillations into oscillatory compartment dynamics~\cite{Smokers2025, Bauermann2026}.

Whether phase separation can itself generate sustained oscillations under stationary chemical driving remains an important question in nonequilibrium soft matter. 
Recent theoretical work has demonstrated that chemically fueled droplets can undergo self-sustained cycles of formation and dissolution without oscillatory reaction kinetics~\cite{Haugerud2026}. 
Complementary experiments on chemically active coacervates confined within synthetic cells revealed remarkably regular oscillations in droplet number, size, and spatial distribution over hundreds of cycles~\cite{sastre2025size}. 
In these experiments, droplets nucleate near the top of the compartment, grow and sediment, fuse at the bottom, and subsequently shrink, suggesting a feedback mechanism involving active size selection and material recycling. 
However, the collective dynamics underlying these observations remain insufficiently understood.
In particular, it is unclear how the coupling between stochastic nucleation, droplet growth, gravitational transport, and fusion selects a reproducible oscillation period and organizes individual droplets into coherent propagating pulses. 
Establishing this connection requires a theoretical description that captures interacting-droplet dynamics while retaining their coupling through the surrounding chemical concentration fields.

Here, we develop a theoretical framework for chemically active emulsions under gravity that quantitatively connects the dynamics of individual droplets to collective oscillations. 
By coupling reaction–diffusion dynamics to stochastic nucleation, droplet growth, sedimentation, fusion, and hydrodynamic transport, we identify the feedback mechanism responsible for the experimentally observed oscillations~\cite{sastre2025size}.
We show that the oscillation period is selected by the interplay between nucleation and sedimentation timescales, leading to distinct nucleation-limited and sedimentation-limited regimes. 
Furthermore, we demonstrate that nucleation kinetics determine whether droplets form temporally dispersed populations or coherent propagating pulses. 
Our results establish how collective oscillatory dynamics can emerge from stationary chemical driving through the coupling of phase separation and transport. 
Beyond explaining a specific experimental realization, the revealed mechanism stimulates the design of autonomous synthetic cells with a programmable internal clock.

\section{Model of an active emulsion under gravity}

To understand what governs the frequency and amplitude of active droplet oscillations under gravity, we propose a coarse-grained theoretical model that quantitatively captures these oscillations.
This model is developed to describe active chemistry in (three-dimensional) emulsions composed of a large number of droplets (from a few to thousands of droplets and more) -- a scale in general difficult to capture by multi-component, stochastic field theories. 
Active chemistry refers to chemical reactions with broken detailed balance of the rates, realized through a fuel chemostat with reaction rate coefficients that differ between the droplet and the surrounding solvent-rich phase. 
The theoretical model is developed for active coacervates composed of charged polyanionic polymers with EDC-fueled active reactions~\cite{sastre2025size}.
However, the phenomenon and the underlying mechanism of gravity-induced oscillations are more general, and the proposed model can be applied to different chemical reactions and phase-separated systems under gravity.

\subsection{Droplet movement, growth and shrinkage}

We propose a discrete droplet description for a large number of droplets that are surrounded by a solvent-rich phase with the droplet number $N(t)$  changing with time $t$. 
Each droplet $i= \{1,2,...,N(t)\}$ is assumed to have a maintained spherical shape, which is a good approximation for mesoscopic condensates below the capillary length and surface tension large enough such that interfacial fluctuations are negligible; see Appendix~\ref{apD:Capillary fluctuations} for estimates showing that most biomolecular condensates and coacervates are consistent with this simplification. 
In this case, each droplet state is characterized by its center of mass coordinate $\textbf{x}_i$ and radius $R_i$. 
Moreover,  mesoscopic droplets undergo Brownian motion with a translational diffusivity $D_i=k_BT/(6\pi \eta R_i)$ obeying the Sutherland-Einstein-Stokes relationship, where $\eta$ is the surrounding solvent-rich phase viscosity. 
On times above the inertial relaxation time (microsecond range for micron-sized objects in water), the droplet mass coordinate $\textbf{x}_i$ obeys an overdamped Langevin equation: 
\begin{equation} \label{eq:BrDy}
     \frac{\text{d}}{\text{d}t} \textbf{x}_i = \frac{\textbf{F}_i}{6\pi\eta R_i} + \textbf{v}(\textbf{x}_i,t) + \boldsymbol{\xi}(\textbf{x}_i,t)\ ,
 \end{equation}
 with uncorrelated, white noise fluctuations the velocity, \begin{equation}\langle \xi_l (\textbf{x}_i,t)\xi_{l'}(\textbf{x}_i,t')\rangle  =  \frac{2k_BT}{6\pi\eta R_i}\delta(t-t')\delta_{l l'}\, , \end{equation} where $l,l' \in \{x,y,z\}$ denotes the spatial components. We assume the noise to be independent for each droplet.

We consider droplets as sparsely distributed meaning that their radii $R_i$ are typically small relative to the interdroplet distances. 
In EDC-fueled coacervates~\cite{sastre2025size}, droplet number density are in the range of $1/(100 \mu m^3)$ with typical droplet size on the micron scale supporting the assumption sparsely distributed droplets. 
This situation allows us to treat the droplets as effective ``points'' in a solvent-rich phase with concentration $c_\alpha(\textbf{x},t)$ for the molecular component $\alpha$ that we refer to as ``background field''.
The different molecules $\alpha$ in the background field can diffuse and are advected with the hydrodynamic flow field velocity $\textbf{v}(\textbf{x},t)$.
This flow field  also advects the point-like droplets in Eq.~\eqref{eq:BrDy} and mediates hydrodynamic interactions between them. 
Moreover, droplets are subject to gravity with the buoyancy force 
\begin{equation} \label{eq:Forceondroplets}
 \textbf{F}_i =- \frac{4\pi R_i^3 \Delta\rho }{3} g \textbf{e}_z \, , 
 \end{equation} 
proportional to the gravitational acceleration $g$, pointing along the negative $z$-direction.
Moreover, the excess mass density \begin{equation*}
 \Delta\rho = \rho^\text{in} - \rho^\text{out} \, , 
 \end{equation*}
 where  $\rho^\text{out}$ and $\rho^\text{in}$  are the mass density of solvent-rich phase and  the droplet, respectively.
 In Ref.~\cite{sastre2025size}, droplets are heavier than the solvent-rich phase with 
$\rho^\text{in} = 1.8 \, g/ cm^{3}$.

The droplets grow or shrink due to diffusive fluxes of molecules across the interface, with the radius $R_i$ corresponding to the interface position for radial symmetric droplets.
For interfaces much thinner than the droplet radii $R_i$,  each droplet's interface velocity 
\begin{equation} \label{eq:Droplet_growth_rate}
    \frac{\text{d}}{\text{d}t} R_i = \frac{j^\text{in}_\alpha(i) - j^\text{out}_\alpha(i)}{c^\text{in}_\alpha(i)-c^\text{out}_\alpha(i)}\, ,
 \end{equation}
follows from local conservation of mass for each component $\alpha$. Here, $j^\text{in/out}_\alpha(i)$ and $c^\text{in/out}_\alpha(i)$ are outward normal diffusive fluxes and concentrations at the droplet interface $R_i$  of droplet $i$ of component $\alpha$. 
Considering the transport equations inside and outside being linearized around their respective phase equilibria (see Ref.~\cite{Weber2019, Bauermann2022, Juelicher2024DropletPhysics} for more information, and Appendix~\ref{apA:Sharp interface limit} for an explicit derivation from coupled Cahn-Hilliard/Allen-Cahn equations), the diffusive fluxes are given in terms of concentration gradients,  
\begin{subequations} \label{eq:fluxes}
    \begin{align}
     j^\text{in}_\alpha &= -D^\text{in}_\alpha (\nabla c_\alpha)^\text{in}\ , \\ j^\text{out}_\alpha &= -D^\text{out}_\alpha (\nabla c_\alpha)^\text{out}\ ,
     \end{align}
\end{subequations}
 where $D^\text{in/out}_\alpha$ are the diffusivities inside the droplet or the surrounding solvent-rich phase. 
 To obtain $j^\text{in/out}_\alpha$ through Eq.~\eqref{eq:fluxes}, we need to determine the concentration profiles inside and outside the droplets.

 For the inside of the spherically symmetric droplet, we use a quasi-static approximation with $\partial_t c_\alpha(r,t)\simeq 0$, where $0<r<R_i$ is the radial position of droplet $i$.
 The quasi-static approximation requires that droplet growth/shrinkage is slow enough such that the concentration profiles inside are stationary at each time point $t$.
 In other words, as the droplet interface moves, grows and or shrinks, the concentrations respond by instantaneously adopting the new steady state profiles inside. At each quasi steady states,  we then solve for the steady state using
\begin{equation} \label{eq:quasi-static}
     0 \simeq D_\alpha^\text{in} \nabla^2 c_\alpha(r)\  + s_\alpha(\{c_\beta(r)\})\, .
 \end{equation} 
 The corresponding eigenfunctions are discussed in Appendix~\ref{apB:Quasi-static droplets}. 
 
Outside droplets, we use a discrete spatial grid over which we define concentration fields as described in Appendix~\ref{appH:num_int_param}. The kinetics of these fields are described in the next subsection. To evaluate fluxes $j_\alpha^\text{out}$ consistently, we interpolate concentrations outside the droplet interface to the discrete grid using the same analytical approach, see Appendix~\ref{apC:Quasi-static layer}. 

 \subsection{Diffusion and chemical reactions in the solvent-rich phase}\label{eq:chem_reactions}

The minimal molecular set of components for active droplets are the precursor $A$ that is converted to the product $B$ through the degradation of a fuel $F$, while deactivation occurs spontaneously:
\begin{subequations}
\begin{align}
   & \text{Activation:  }   \ce{\textit{A} \xrightarrow{\text{fuel-driven:} \, \, \, k_\text{act} c_F} \textit{B}}  \, , 
   \label{eq:forward_react}\\ 
   & \text{Deactivation:  } \ce{\textit{B} \xrightarrow{\text{spontaneous:} \, \, \, k_\text{deact}} \textit{A}} \, , 
\label{eq:backward_react}
\end{align}
\end{subequations}
where $k_\text{deact}$ and $k_\text{act}$ denote the deactivation and activation rate constants and $c_\text{F}$ is the fuel concentration controlling the activation pathway. 
In Ref.~\cite{sastre2025size}, the precursor $A$ is a peptide that changes its charge state, leading to $B$ that has an enhanced phase separation propensity.
This activation is mediated by the fuel EDC that turns over to EDU.
For this reaction, $s_A=-s_B=:s$ due to mass conservation, and fuel-driven, active reactions admit a simple mass action law-like kinetics (see Ref.~\cite{Weber2019} for  discussions on broken detailed balance of the rates):
\begin{align} \label{eq:reaction-rate}
    s &= k_\text{act}c_Fc_A - k_\text{deact}c_B\ , \\
s_F &= -k_\text{act}c_Fc_A  \, . 
\end{align}
Moreover, the concentrations are denoted as  $c_A$ for precursor, product $c_B$ for product, and  $c_F$ for fuel. For the background field, we therefore solve the corresponding linear transport equations corresponding to reaction-diffusion equations, coupled between the different molecular components $\alpha$:
\begin{align} 
\nonumber
     \partial_t c_\alpha(\textbf{x},t) &= D_\alpha^\text{out} \nabla^2 c_\alpha(\textbf{x},t) \  - \textbf{v}(\textbf{x},t)\cdot \nabla c_\alpha(\textbf{x},t)
     \\
     &\quad+ s_\alpha(\{c_\beta(\textbf{x},t)\}) - r_\alpha(\textbf{x},t)\, , \,    \label{eq:reaction-diffusion}
 \end{align}
 where the local sink/source terms corresponding to droplet growth/shrinkage) reads
 \begin{equation} \label{eq:source/sink}
    r_{\alpha}(\textbf{x},t) = \sum_i \delta(\textbf{x}-\textbf{x}_i)(c_{\alpha}^\text{in} - c_{\alpha}^\text{out}) \, 4\pi R_i^2 \frac{\text{d}}{\text{d}t} R_i \, .
\end{equation} 
Note that the  reaction flux
 $s_\alpha(\{c_\beta(\textbf{x},t)\})$ depend on the set of components $\{c_\beta\}$, mediating the coupling between different components $\alpha$.

Like in the experimental set-up, the fuel in our system is maintained constant at the system boundary, and is consumed inside the system at all times. This creates a flux of the fuel from the fuel reservoir into the system, which maintains the system out of thermodynamic equilibrium. The expression of chemical rates in Eq.~\eqref{eq:reaction-rate} is obtained by expanding non-linear chemical fluxes, governed by chemical potentials as explained in Ref.~\cite{Weber2019}, around local phase equilibrium concentrations. 
We solve Eq.~\eqref{eq:reaction-diffusion} for the fuel concentration $c_F$ outside droplets. We see that the fuel diffuses fast enough such that the fuel outside is very close to the value at the system boundary $c^\text{out}_F\simeq c_F|_b$.
For simplicity, we assume that the fuel concentration does not change inside droplets over the length scale of the droplet radius, and is simply $c^\text{in}_F=P_F \, c^\text{out}_F$, where $P_F$ is the fuel partitioning coefficient.

\subsection{Nucleation of droplets} 

The droplet number $N(t)$ in the system varies with time, because droplets nucleate and fuse. 
The rate of nucleation is largely governed by the supersaturation field of component $\alpha$,
\begin{equation} \label{eq:Supersaturation}
\epsilon_\alpha(\textbf{x},t) = \frac{c_\alpha(\textbf{x},t) - c_\alpha^{\text{out}}}{c_\alpha^\text{out}}\ ,
\end{equation}
 where $c_\alpha(\textbf{x},t)$ is the local background concentration field at a point $\textbf{x}$ and time $t$. Moreover, $c_\alpha^\text{out}$ is the threshold (binodal) concentration dictated by the system's phase diagram.
 See Ref.~\cite{sastre2025size} for the experimental phase diagram and a fit based on a Flory-Huggins free energy density $f$.  
For our minimal model of active droplets, we have two non-solvent components $A$ and $B$.
Thus, the effective supersaturation field
 \begin{equation} \label{eq:epsilon-ternary}
     \epsilon = \sum_\alpha \sum_\beta  \frac{c_\beta^\text{out}(c_\alpha^\text{in}-c_\alpha^\text{out})f''_{\alpha\beta}}{c_B^\text{out}(c_B^\text{in}-c_B^\text{out})f''_{B}} \epsilon_\beta(\textbf{x},t) \ ,
 \end{equation} 
 depends on two concentration fields $c_A(\textbf{x},t)$ and $c_B(\textbf{x},t)$ (derivation see Appendix~\ref{apF:Nuc_rate}). 
 Here,  
$f''_{\alpha\beta}$ is the Hessian of the free energy density that is fixed by the experimental phase diagram. 
Moreover, $c_{\alpha}^\text{in/out}$ are the phase equilibrium concentrations inside and outside the droplet. For simplicity in the simulations, we consider only the diagonal terms of the Hessian $f''_{\alpha \beta}$, which simplifies the expression for $\epsilon$ to \begin{equation}\label{eq:epsilon-simplified}
    \epsilon =   \frac{c_A^\text{out}(c_A^\text{in}-c_A^\text{out})f''_{A}}{c_B^\text{out}(c_B^\text{in}-c_B^\text{out})f''_{B}} \epsilon_A(\textbf{x},t) + \epsilon_B(\textbf{x},t)\ .
\end{equation}
Given the effective supersaturation field  $\epsilon(\textbf{x},t)$, the average rate of nucleation of new droplets at each position $\textbf{x}$ is approximated through Kramers' theory~\cite{hanggi1990reaction}:  \begin{equation} \label{eq:nucleation-rate}
    \tau^\text{-1}_\text{nuc}(\textbf{x},t) = K\exp{\left(-\frac{16 \pi \gamma^3 }{3 (f''_Bc_B^\text{out}(c_B^\text{in}-c_B^\text{out}))^2 k_BT \epsilon(\textbf{x},t)^2 }\right)}\, ,
\end{equation} where $\tau_\text{nuc}$ denotes the expectation value of the time between two successive nucleation events. We will henceforth refer to this as the nucleation time.
To simulate each individual nucleation event, we use the kinetic Monte Carlo algorithm (details see Appendix~\ref{appH:num_int_param}).

\subsection{Micro-reactor geometry and dynamics of the active, bottom-wetted layer} 

Droplets nucleate, grow and sediment following Eqs.~\eqref{eq:BrDy}, \eqref{eq:Forceondroplets}, and \eqref{eq:Droplet_growth_rate}.
At the bottom, gravity induces fusion. For simplicity and to avoid modeling geometry-related details (e.g., sliding motion on a spherical bottom surface), we consider a simplified {micro-reactor} geometry compared to the spherical one in Ref.~\cite{sastre2025size}, i.e., a rectangular box with gravity acting along $z$ with a domain height $H$, while  the $x,y$-directions have the dimensions $L$ with mixed boundary conditions; Neumann for components $A$ and $B$ and Dirichlet for fuel $F$.

In such a rectangular geometry, gravity-induced fusion leads to a wetted droplet layer at the bottom with a dynamic height $h(t)$. 
The wetted layer grows or shrinks at a rate given by 
\begin{equation} \label{eq:height-growth-rate}
    \frac{\text{d}}{\text{d}t} h(t) = \frac{j^\text{in}_\alpha-j^\text{out}_\alpha}{c^\text{in}_\alpha-c^\text{out}_\alpha} \, , 
\end{equation}
where, as in Eq.~\eqref{eq:Droplet_growth_rate},  
$j^\text{in/out}_\alpha$ are the diffusive fluxes and $c^\text{in/out}_\alpha$ the  concentrations of component $\alpha$ inside/outside at the droplet interface.
Similar to the sedimenting active droplets, the active, bottom-wetted layer has a stable fixed point $h_\text{stat}$ above which $h(t)$ shrinks, while below the layer grows; see Fig.~\ref{fig:fig2}(f). The derivation of concentration profiles for the wetted layer are described in Appendix~\ref{apC:Quasi-static layer}.

\subsection{Hydrodynamic interactions between active  sedimenting droplets}  

As droplets grow and sediment, they drag the surrounding fluid with them, generating flow in the system. 
For micron-sized droplets in the small Reynolds number regime, the flow velocity $\textbf{v}(\textbf{x},t)$ follows the quasi-static  Stokes equation
 \begin{align} \label{eq:Stokes}
    -\nabla P(\textbf{x}) + \eta \nabla^2 \textbf{v}(\textbf{x},t) + \sum_i \textbf{F}_i \delta(\textbf{x}-\textbf{x}_i) = 0 
\end{align}
with droplets creating local, point forces $F_i$ (Eq.~\eqref{eq:Forceondroplets}). Here $\eta$ is the viscosity of the solvent-rich phase, same as defined in Eq.~\eqref{eq:BrDy}.
Since aqueous solvents are almost 
incompressible, we use $\nabla \cdot \textbf{v}(\textbf{x},t) = 0 $ at all times $t$.

\subsection{Model integration and parameter choices} 

We now have the full set of equations required to describe the evolution of the system containing multiple droplets in time. 
In the following, we integrate Eqs.~\eqref{eq:BrDy},~\eqref{eq:Droplet_growth_rate},~\eqref{eq:quasi-static},~\eqref{eq:height-growth-rate}  in time. In addition, the droplet number changes stochastically, at an average rate given by Eq.~\eqref{eq:nucleation-rate}, and due to fusion. 
At every time step, the stationary solutions of Eq.~\eqref{eq:quasi-static} give the internal concentration profiles inside droplets necessary to integrate Eq.~\eqref{eq:Droplet_growth_rate}, and that of Eq.~\eqref{eq:Stokes} give the velocity field of the fluid surrounding droplets. For clarity, we summarize the underlying model assumptions in Appendix~\ref{apG:model summary}.

The numerical scheme used to solve for the model in time is given in Appendix~\ref{appH:num_int_param} together with a list of all parameters and their values. 
All parameters except the nucleation rate constant $K$ are obtained or estimated from experimental measurements, and are therefore realistic. 
The thermodynamic parameters such as molecular volumes and interaction parameters are obtained from fitting the experimental phase diagram; see Ref.~\cite{sastre2025size}. 
The diffusivities $D$, reaction rate coefficients $k_\text{deact}$ and $k_\text{act}$, and the density difference $\Delta \rho$ were also measured experimentally. 
The nucleation rate constant $K$ in Eq.~\eqref{eq:nucleation-rate} is not directly measurable from experiments. 
Its value is set by the physicochemical properties of the system boundary, which is an oil-aqueous phase interface. 

Since $K$ changes the nucleation time $\tau_\text{nuc}$ (Eq.~\eqref{eq:nucleation-rate}), which in turn depends on the effective supersaturation $\epsilon$, we will see that $K$ changes the width of droplet pulses.

\section{Results and discussion}
\begin{figure*}[tb]
    \includegraphics[width=\linewidth]{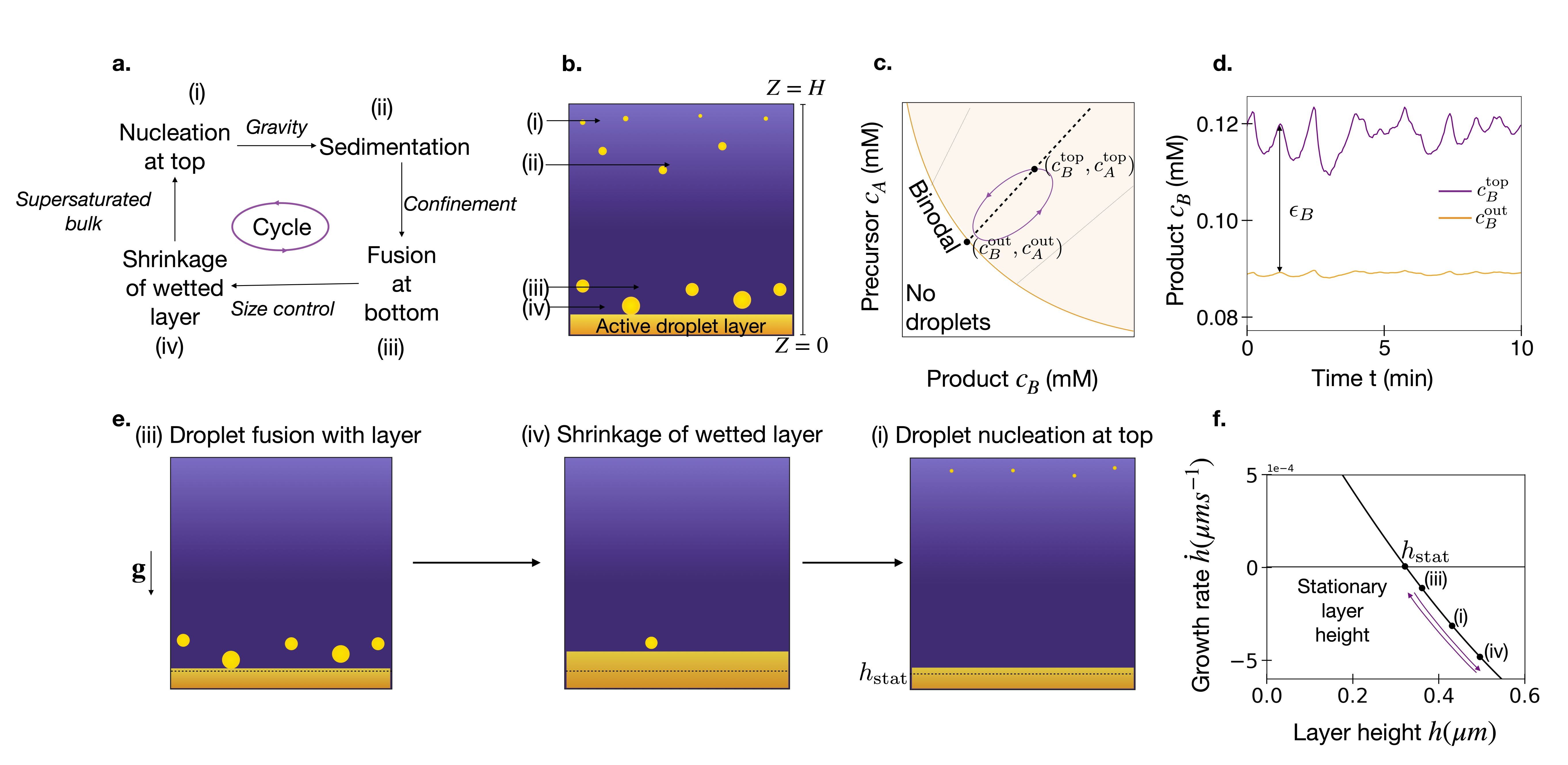}
    \caption{
    \textbf{Mechanism of oscillating chemically-fueled droplets: four stages of the cycle.}
    (a,b) The shrinkage of the bottom active droplet layer raises the supersaturation at the top, leading to nucleation of new droplets and their subsequent growth (i).
    This growth in turn decreases the supersaturation (c,d), arresting nucleation of further droplets.  
    When the freshly nucleated droplets grow big enough, they start sedimenting together as a pulse, finally causing fusion with the bottom active droplet layer.   
    (e,f) The resulting larger layer height causes an efflux of droplet material through the established size control mechanism of an active droplets~\cite{Zwicker2015} related to the deactivation of the droplet component $B$ inside to the less phase-separating $A$. 
    The negative feedback of droplet growth  and the positive feedback of bottom layer shrinkage on nucleation at the top is evidenced in oscillations of $A$ and $B$ concentration in the bulk (c), and oscillations of the supersaturation $\epsilon_B$ of the phase separating components at the top surface (d). Sub-figures (b), (c), (e) are illustrative images, while (d) and (f) represent simulation results with fuel concentration $c_F = 3.3$mM, conserved quantity $\psi = 10$mM and {micro-reactor} height $H=150\mu m$.}
    
    \label{fig:fig2}
\end{figure*}

\begin{figure*}[tb]
    \includegraphics[width=\linewidth]{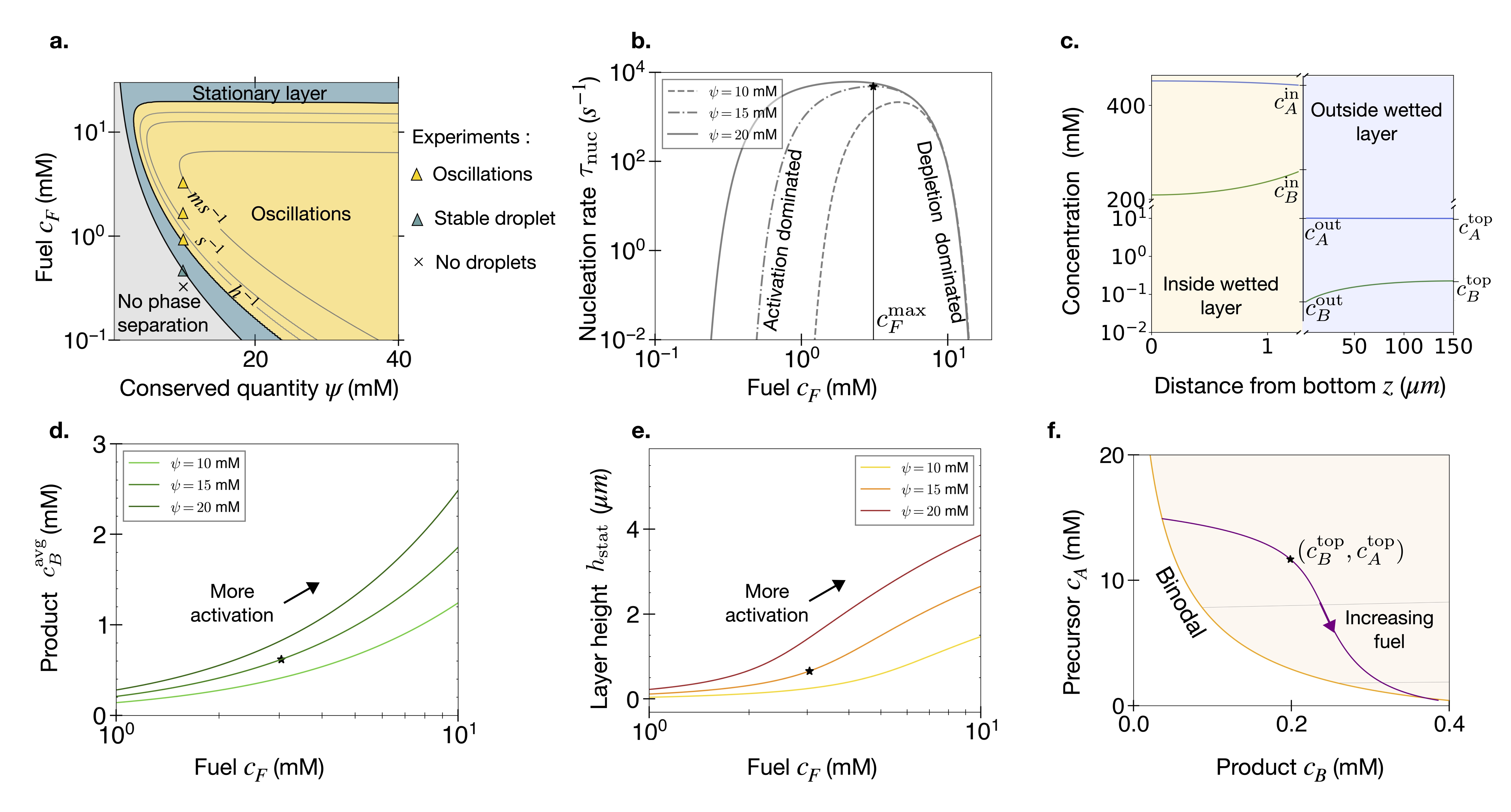}
    \caption{\textbf{
    Gravity-induced active droplet oscillations are regulated by fuel $c_F$ and total droplet material $\psi$.}
    (a) State diagram of gravity-induced active droplet oscillations: 
    A high fuel concentration drives nearly complete conversion from $A$ to $B$, reducing efflux and thereby weakening the size-selection mechanism. Thus, oscillations vanish, leading to a stationary active droplet layer (blue).
    Moreover, oscillations (yellow) require sufficient deviations from the binodal to nucleate droplets at the top.  
(b) The rate of droplet nucleation shows a maximum at $c_F=c_F^\text{max}$, which arises from an activation/depletion-dominated regime:
    upon increasing $c_F$, more precursor ($A$) gets activated to product ($B$), enhancing total concentration of activated product $c_B^\text{avg}$ in the system (d). 
     The layer height $h_\text{stat}$ strongly governs the composition at the top of the {micro-reactor} ($c_A^\text{top},c_B^\text{top}$) (c). The precursor $A$ typically has smaller gradients than the product because of differences in diffusivity. 
    The stationary height of the wetted droplet layer also increases with increasing fuel (e), with the activated product partitioning into the wetted layer.
    Therefore, not all of the activated product remains in the outside phase, forcing the composition at the top of the {micro-reactor} ($c_A^\text{top}$, $c_B^\text{top}$) to follow a non-linear trajectory with increasing fuel (f). 
    At very high fuel, the composition ($c_A^\text{top}$, $c_B^\text{top}$) touches the binodal line of the phase diagram, meaning $\epsilon =0$, which completely arrests nucleation. 
    The competition between fuel-induced activation and depletion due to layer growth (d,e,f) creates an optimum fuel concentration where supersaturation, and hence nucleation rate, is maximal (b). The {micro-reactor} height was fixed to $H = 150 \mu m$ for all sub-figures from (a) to (f).
   } 
    \label{fig:fig3}
\end{figure*}

\begin{figure*}[tb]
    \includegraphics[width=\linewidth]{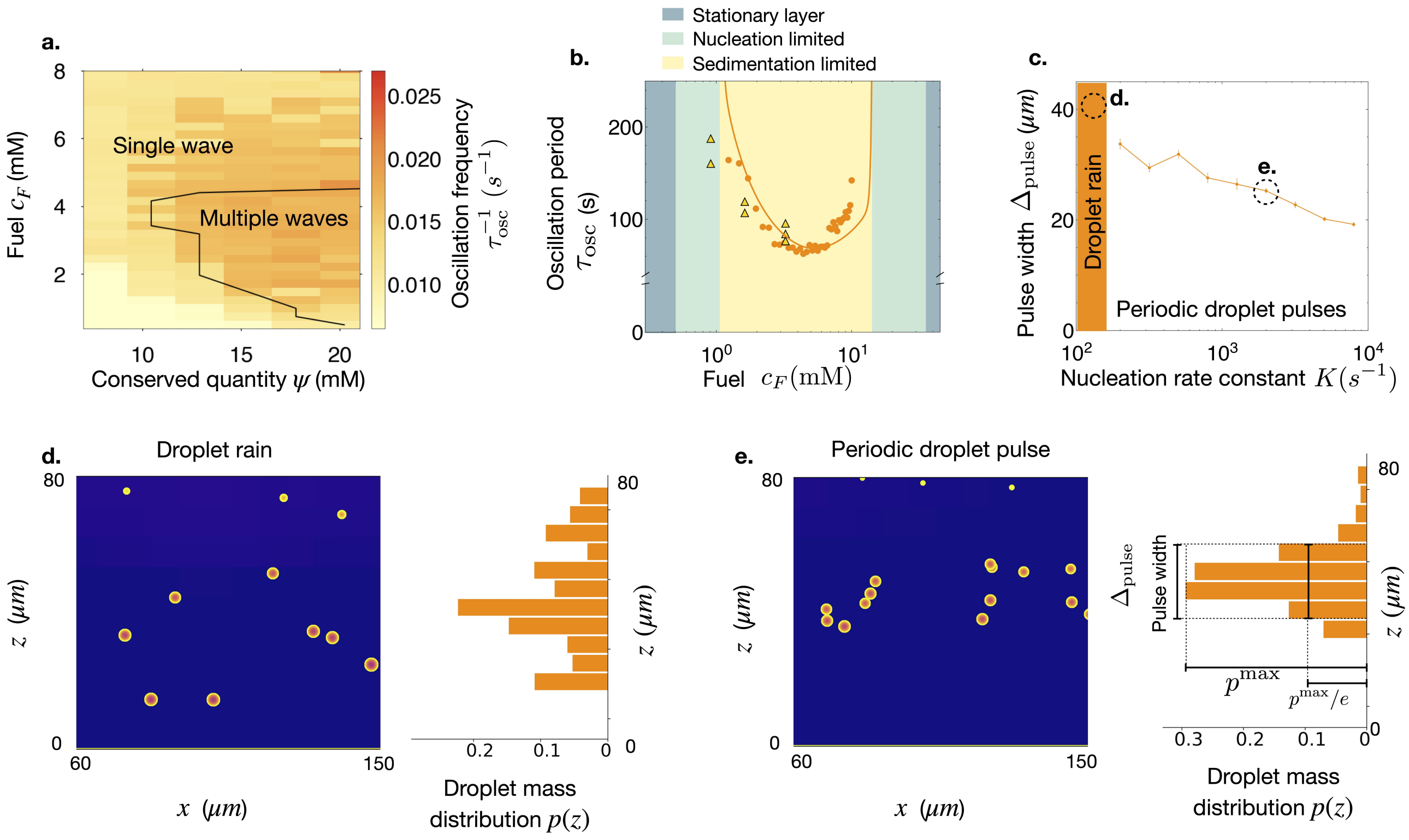}
    \caption{ \textbf{
    Transition from a single droplet pulse to multiple waves, or continuous droplet rain.}
    (a) There is a transition to multiple waves, where droplets nucleate one after the other before the previous waves reach the bottom.
    Multiple waves are favored for larger conserved quantity $\psi$ and intermittent fuel concentrations $c_F$. 
    Increasing $c_F$ lowers supersaturation(Fig.~\ref{fig:fig3}f), suppressing the formation of multiple waves.
    The transition from single pulses to multiple waves is defined by the points where $\tau_\text{sed}/\tau_\text{nuc} = 3.2$; see the black line. 
    (b) The oscillation period $\tau_\text{osc}$ shows a local minimum with $c_F$ in a regime where nucleation is fast, and the period $\tau_\text{osc}$ is limited only by the sedimentation time $\tau_\text{sed}$. 
     The oscillation period obtained in simulations (orange dots) results from the nucleation and sedimentation  timescales, $\tau_\text{nuc}$ and $\tau_\text{sed}$, and matches the experimentally observed period (yellow triangles).
     The combined timescale $(\tau_\text{nuc}+\tau_\text{sed})$ (Eq.~\eqref{eq:nucleation-rate}) from analytical estimates (solid orange line) captures the oscillation period in simulations (orange dots). 
     For fuel concentration $0.3<c_F<1$mM and $10<c_F<15$ mM, the nucleation timescale $\tau_\text{nuc}$ is rate limiting. 
     For intermediate fuel concentration $1<c_F<10$mM we see that nucleation is much faster, and the sedimentation $\tau_\text{sed}$ is the limiting timescale.
    (c,d,e) As we increase the nucleation rate constant $K$, the oscillations transition from 
    droplet rain to 
    droplet pulses.
    Droplet rain corresponds to sedimenting  droplets that are approximately distributed uniformly along $Z$;
 see snapshots and histograms of droplet distribution along the $z-$direction with droplet rain (d) and droplet pulse (e).
Droplet rain originates because nucleation at the top occurs without well-defined breaks.
The snapshots show $x\in[60,150]\ \mu m$ of the micro-reactor. 
    Moreover, the width of droplet pulses $\Delta_\text{pulse}$ (c) decreases steadily with $K$ because droplet nucleation becomes more synchronized, so pulses become narrower.}
    \label{fig:fig4}
\end{figure*}

In this section, we use our discrete droplet model Eqs.~\eqref{eq:BrDy}-~\eqref{eq:Stokes} to gain insight into the collective emergence of oscillations, unraveling the prerequisites of oscillations in a chemically active emulsion in gravity. 
We discuss the intrinsic timescales that set the oscillation frequency and its variation with our control parameters. 
The oscillations are characterized by traveling droplet pulses in space and time that reappear via nucleation in a periodic fashion at the top of the system.
We investigate how the width of these pulses can be tuned, marking a transition from compact, well-defined pulses to a droplet rain.

\subsection{Emergence and perpetuation of oscillations in a multi-droplet system}

The three key requirements for oscillations of active droplets in external fields such as gravity:

\begin{itemize}
    \item[\textit{(1)}] An \textit{active size control mechanism} by which droplets above a certain size shrink and release material in the surrounding phase.
    
    \item[\textit{(2)}] A \textit{directed fusion mechanism}, for example, an external field like gravity, that advects droplets toward certain spatial domains, enforcing their fusion and growth above their stationary size locally.
    
    \item[\textit{(3)}] A supersaturated background phase capable of \textit{nucleating  droplets at the system boundary} which is fed by the shrinking bottom active droplet layer.  
\end{itemize}

\textit{(1) Active size-control mechanism.} 
Active droplets described in the previous section have been shown to exhibit a stationary size above which they shrink~\cite{Zwicker2015}. 
In particular, the more strongly phase separating component $B$ (product of the forward reaction~\eqref{eq:forward_react}) is deactivated inside the active droplets, and activated outside active through the hydrolysis of a fuel component $F$ to waste (Fig.~\ref{fig:fig1}(a,d)). 
This chemical flux creates a diffusive efflux of the less phase-separating component $A$, which competes with droplet coarsening in the supersaturated background field. 
Simply put, product deactivation scales with droplet volume $\sim R^3$, while material influx scales with droplet radius $R$~\cite{Zwicker2015}.
Thus, droplets smaller than a certain size $R_\text{stat}$ are influx-dominated, leading to growth. 
Droplets larger than $R_\text{stat}$  shrink. 
Various aspects of this principle have been theoretically discussed in detail in several works~\cite{Zwicker2015,wurtz2018chemical, Bauermann2025}, and it was shown to occur for synthetic coacervate~\cite{sastre2025size}. 
Similar to droplets, the wetted layer that forms when the droplets in our model system fuse at the bottom has a stationary height $h_\text{stat}$ above which it shrinks (Fig.~\ref{fig:fig2}(e,f)). 

\textit{(2) Directed fusion mechanism.} 
At steady state, the emulsion becomes monodisperse if the droplets remain suspended in the solution~\cite{Zwicker2015, Bauermann2025}.
However, in the presence of an external gradient or field, such as gravity, droplets with higher mass density than their environment sediment to the bottom of the {micro-reactor}, where they undergo directed fusion, causing the bottom layer to grow (or vice versa). 
If the bottom active droplet layer becomes larger than the stationary size $h_\text{stat}$, it shrinks (see \textit{1}). 
This efflux drives upward diffusion of deactivated droplet material, raising supersaturation at the top again. 

\textit{(3) Nucleating droplets at the top.} Once the effective supersaturation $\epsilon$ becomes large enough due to the shrinking bottom droplet layer, it triggers nucleation of new droplets and the cycle continues.
If the stationary size $h_\text{stat}$  is reached and the supersaturation $\epsilon\le0$, the system remains stationary thereafter.
The parameters that determine whether $\epsilon>0$ while the bottom layer is still shrinking are the fuel concentration $c_F$ and the amount of material $\psi$ in the system. 
In the next subsections, we discuss how these two control parameters determine the onset and frequency of oscillations.

\subsection{Droplet nucleation causes negative  feedback leading to oscillation onset}

In accordance with the previous discussion, the gravity-induced oscillations of active droplets are characterized by four stages (Fig.~\ref{fig:fig2}(a,b)):
\begin{itemize}
    \item[(i)] droplet nucleation at the top surface,
    \item[(ii)] growth and sedimentation of nucleated droplets,
    \item[(iii)]  fusion at the bottom driven by gravity, and 
    \item[(iv)] shrinkage of the bottom layer.
\end{itemize}

\textit{Nucleation and negative feedback.} For the supersaturation $\epsilon > 0$, the rate of nucleation $\tau_\text{nuc}^{-1}$ is non-zero, and with increasing $\epsilon$, $\tau_\text{nuc}^{-1}$ sharply increases. 
Upon nucleation, the nucleated droplets take up material from the background phase and grow.  
Droplet growth locally decreases $c_A$ and $c_B$ in the vicinity of droplets, thus lowering the effective supersaturation $\epsilon$ (Fig.~\ref{fig:fig2}(c,d)).  
This process provides negative feedback that arrests further nucleation by locally reducing the effective supersaturation $\epsilon$ and thus strongly decreasing the nucleation rate. 
The resulting nucleated pulse of droplets continues to grow, depleting $A$ and $B$ further until nucleation is practically arrested. 

\textit{Site of nucleation.} In the theoretical model, nucleation of droplets occurs primarily in the top part of the {micro-reactor} ($z \simeq H$). 
Nucleation at the top is also seen clearly in the experiments (Fig.~\ref{fig:fig1}(b)). 
Using our model we can explain this behavior by the concentration profiles $c_A$ and $c_B$ outside the bottom layer. 
The concentration at the bottom layer interface ($z = h$) is always at phase equilibrium ($c_A^\text{out}$ and $c_B^\text{out}$) for thin interfaces, see Fig.~\ref{fig:fig3}(c). 
Therefore, the composition of the solvent-rich phase near the layer interface ($z \ge h$) is closer to the dilute binodal or the saturation limit. 
In other words, supersaturation $\epsilon$ is higher for points further from the bottom layer interface, resulting in a higher nucleation rate $\tau_\text{nuc}^{-1}$ at the top. 
Once droplets nucleate near the top, they naturally inhibit any nucleation in other parts of the {micro-reactor}.  

\textit{Sedimentation and fusion.} As a droplet of size $R$ grow, its terminal velocity through the background fluid increases. 
The gravitational P\'eclet number (Pe) associated with droplet sedimentation across the {micro-reactor} increases as droplets grow. It is given by $\text{Pe} = v_\text{sed}R/D_\text{drop}$, where $D_\text{drop} = k_BT / 6\pi \eta R$ is the droplet diffusivity.
When it becomes large enough ($\text{Pe} > 1$), droplets start to sediment (Fig.~\ref{fig:fig1}(e)). 
Droplets then fuse at the bottom so that the bottom layer becomes larger than its stationary size $h > h_\text{stat}$, and shrinks through an efflux of material into the background phase.
This mechanism  closes the series of events to a cycle, finally leading to oscillations in droplet volume $V_d$, supersaturation $\epsilon$ and the size of the bottom layer $h$ (Fig.~\ref{fig:fig1}(f),~\ref{fig:fig2}(d) and ~\ref{fig:fig2}(f)). 

We calculate the oscillation frequency $\tau_\text{osc}^{-1}$ by Fourier transform of the time series of the droplet volume in the uppermost part of the micro-reactor, with $\tau_\text{osc}^{-1}$ corresponding to the maximum of the power spectrum. The  oscillation frequency $\tau_\text{osc}^{-1}$ with fuel concentration $c_F$ and conserved quantity $\psi$ is shown in Fig.~\ref{fig:fig4}(a).
We see that the frequency $\tau_\text{osc}^{-1}$ shows a maximum value around $c_F=3$mM and $\psi=18$mM. 
Moreover, there is a transition of single sedimenting droplet wave per cycle to multiple droplet waves; see Fig~\ref{fig:fig4}(a). The boundary separating the single wave regime and the multi-wave regime in Fig.~\ref{fig:fig4}(a) is governed by the supersaturation level near the top of the micro-reactor. Since supersaturation level is lower for very high $c_F$ as shown in Fig.~\ref{fig:fig3}(f), a new wave cannot nucleate until the first wave sediments and fuses with the layer. Typically, for fuel concentration less than $1$mM and greater than $15$mM, we see that there are no oscillations because droplet nucleation is completely arrested; see Fig~\ref{fig:fig3}(a,b) for analytical estimates and Fig~\ref{fig:fig4}(a) for simulation results.

To investigate this dependence further, we looked at the key time-scales in the system, and how they are affected by $c_F$ and $\psi$.

\subsection{Nucleation and sedimentation timescales govern oscillation frequency}

Here we explain that the oscillation frequency is governed by a series of events: 
Shrinkage of the wetted droplet layer at the bottom increases the effective supersaturation $\epsilon$ in the solvent-rich phase, triggering nucleation. 
Once nucleation is arrested, the next nucleation cycle starts only after the previous droplet pulse fuses at the bottom. 
This series of events implies for the nucleation time $\tau_\text{nuc}$, and the growth and sedimentation time $\tau_\text{sed}$ that the oscillation time can approximately be written \begin{equation} \label{eq:oscillation period}\tau_\text{osc} \simeq \tau_\text{nuc} + \tau_\text{sed}\,.
\end{equation} 
To scrutinize this phenomenological relation, we estimate the nucleation rate $\tau_\text{nuc}^{-1}$ and the sedimentation rate $\tau_\text{sed}^{-1}$ analytically and compare the 
effective timescale $\tau_\text{nuc} + \tau_\text{sed}$
to the oscillation period $\tau_\text{osc}$ obtained from simulations and experimental data from Ref.~\cite{sastre2025size}.

\textit{Estimate of the nucleation rate.} The nucleation rate $\tau_\text{nuc}^{-1}$ depends on the effective supersaturation $\epsilon$ (Eq.~\eqref{eq:nucleation-rate}). 
Though the supersaturation field $\epsilon$ in general varies in space and time, we focus on its position dependence along the axis of gravity, $\epsilon(z=H,t)$, since nucleation dominates at the top. 
To see how the effective supersaturation $\epsilon$ near $z=H$ varies with $c_F$ and $\psi$, we use a reference case, where the bottom layer is at its stationary height and nucleation has not yet occurred. 
In this setup, we determine the composition at the top by solving for the stationary concentration profiles $c_A(z)$, $c_B(z)$, inside the layer $0<z<h$ and outside $h<z<H$. 
This gives us $\epsilon(z=H)$ when the layer is at $h=h_\text{stat}$, for different values of $c_F$ and $\psi$. 
We then map $\epsilon$ to the nucleation rate $\tau_\text{nuc}^{-1}$ (Fig.~\ref{fig:fig3}(a)). 

\textit{Maximal nucleation rate with fuel concentration.} For a given concentration of conserved quantity $\psi$, the nucleation rate $\tau_\text{nuc}^{-1}(c_F)$ shows a non-monotonic behavior with a local maximum for a certain fuel concentration $c_F=c_F^\text{max}$ (Fig.~\ref{fig:fig3}(b)). 
The initial increase with a fuel concentration arises from more of the droplet component $A$ gets activated to $B$ (Fig.~\ref{fig:fig3}(d)), which pushes the composition of the solvent phase $(c_A,c_B)$ deeper into the binodal region, enhancing supersaturation $\epsilon$.  
However, with increasing $c_F$, the stationary height of the wetted layer $h_\text{stat}$ also increases (Fig.~\ref{fig:fig3}(e)).
This increase in the stationary active layer height $h_\text{stat}$, due to material conservation, pulls the composition of the solvent-rich phase closer to the dilute binodal (Fig.~\ref{fig:fig3}(f)).
The competition between the two processes -- fuel based activation and partitioning of molecules into the layer phase -- lead to a trajectory of the composition of the solvent-rich phase $(c_A,c_B)$ with increasing $c_F$ (Fig.~\ref{fig:fig3}(f)). 
It is clear that for an intermediate fuel value, the distance of the corresponding point $(c_A,c_B)$ in the composition space from the binodal is maximal, implying a maximum in the effective supersaturation $\epsilon$ and in turn maximum in the nucleation rate $\tau_\text{nuc}^{-1}$. 

\textit{Estimate of the sedimentation rate.} Next, we estimate the sedimentation rate for a spherical droplet of radius $R$. The terminal sedimentation speed of an isolated colloidal particle without hydrodynamic interactions is~\cite{happel2012low} \begin{equation} \label{eq:sed_speed}
    v_\text{sed} = \frac{2}{9} \frac{\Delta \rho g R^2}{\eta}\ ,
\end{equation}
where $\Delta\rho$ is the density difference between droplets and solvent-rich phase, $g$ is acceleration due to gravity and $\eta$ is the viscosity of the solvent-rich phase. 

This expression for sedimentation speed offers a good approximation for our droplets, even though they are liquid droplets, because the Weber number \textit{We} associated with our sedimenting droplets is of the order $10^{-11}$, which means they remain almost perfectly spherical as they sediment (see Appendix~\ref{apE: dimensionless numbers}). 
Moreover, the experimentally estimated values of diffusivities inside such droplets suggest that the droplet environment is much more viscous compared to the solvent phase, roughly by a factor $10^4$ (see Appendix~\ref{appH:num_int_param} for parameters table). 
This allows us to neglect flows inside droplets and assume that our droplets sediment like colloidal spheres. 
Given the above, the $z-$coordinate of an isolated sedimenting droplet is given as \begin{equation} \label{eq:sed_time}
   z(t) =  H - \int_0^t v_\text{sed}(R(t)) dt \ ,
\end{equation} where $H$ is the total height of the micro-reactor.
The time $t$ at which the droplet meets the wetted layer such that $z = h$, where $h$ is the layer height, is then the sedimentation time $\tau_\text{sed}$. 
For simplicity of the estimate, we assume that the droplets grow at a roughly constant rate as they sediment. 
Given the droplet radius $R(t)$, the sedimentation speed $v_\text{sed}$ can be approximated by Eq.~\eqref{eq:sed_speed}. We can then estimate the total sedimentation time using Eq.~\eqref{eq:sed_time}.

\textit{Rate-limiting process of oscillations.} 
When we compare the phenomenological ansatz for the oscillation period,  $\tau_\text{nuc} + \tau_\text{sed}$ (Eq.~\eqref{eq:oscillation period}) as a function of fuel concentration $c_F$, to the oscillation frequency $\tau_\text{osc}$ obtained from the multi-droplet simulations, we find good agreement, despite simplifying assumptions in the ansatz such as a constant growth rate during sedimentation and using terminal sedimentation speed for an isolated droplet (orange dots, Fig.~\ref{fig:fig4}(b)). 
We also compare the oscillation period with the experimental data in Ref.~\cite{sastre2025size} (triangles), which also agree well, confirming the model's mechanism for gravity-induced active droplet oscillations in the experiment.
Furthermore, the theoretical model reveals that there are three distinct regions upon varying the fuel concentration $c_F$: 
two nucleation-limited regions enclosed by a sedimentation-limited region (Fig.~\ref{fig:fig4}(b)). 
In the nucleation-limited regions, nucleation is a very slow process because of lower supersaturation $\epsilon$, and the estimated nucleation time $\tau_\text{nuc}\gg\tau_\text{sed}$. 
Therefore, the oscillation period $\tau_\text{osc}\simeq \tau_\text{nuc}$. 
In the sedimentation-limited regime, nucleation is much faster once the wetted layer shrinks, but the nucleated droplets must sediment and fuse at the bottom to trigger a new pulse, making sedimentation rate-limiting. 
In this case, $\tau_\text{nuc} \ll \tau_\text{sed}$, $\tau_\text{osc}\simeq \tau_\text{sed}$. 

\subsection{Nucleation rate tunes droplet pulse width}
\label{subsec:pulsewidth}
Droplet nucleation events are stochastic; thermal fluctuations in concentration fields lead to local nucleation events. 
In our discrete droplet model, we assumed that nucleation is a memory-less stochastic process and thus we used a grid-based kinetic Monte Carlo method to simulate nucleation events (details see Appendix~\ref{appH:num_int_param}).  
In other words, the time interval between two consecutive nucleation events is Poisson distributed with a mean equal to $\tau_\text{nuc}$. 
After nucleation, each droplet starts growing and sedimenting, creating a spatial separation between itself and droplets that nucleate later. If nucleation is fast enough, we see a collection of droplets that nucleate at the start of every cycle and sediments collectively. We refer to this pattern as a ``droplet pulse". For slower or prolonged nucleation, we see a rain of droplets mostly without gaps, since droplets reach the bottom before supersaturation $\epsilon$ drops below threshold. This is shown in Fig~\ref{fig:fig4}(d),(e) and also in Fig~\ref{fig:fig4}(a).
In a regime where we see droplet pulses, we typically observe one pulse, except when a freshly nucleated pulse is at the top and the tail of the bottom pulse fuses with the active droplet layer Fig~\ref{fig:fig4}(e). In simulations, we use the droplet number probability density along the height of the {micro-reactor} $p(z,t)$ to define the width of these pulses $\Delta_\text{pulse}$. The spread of $p(z,t)$ around its peak $p^\text{max}$ gives $\Delta_\text{pulse}$, refer to Fig~\ref{fig:fig4}(e) and Appendix~\ref{appH:num_int_param} for more details.

\textit{Analytical ansatz} 
We motivate the dependence of pulse width $\Delta_\text{pulse}$ on the nucleation rate with the following analytical argument.

The average pulse width can be approximately expressed in terms of droplet sedimentation speed $v_\text{sed}$ and the total duration $\Delta t$ over which droplet nucleation happens for a given pulse 
\begin{equation} 
    \langle \Delta_\text{pulse} \rangle \simeq  v_\text{sed}  \langle\Delta t\rangle\, ,
\end{equation}
where the $\langle \Delta_\text{pulse} \rangle$ is the pulse width averaged over multiple cycles, and $\Delta t$ is the time interval between nucleation of the first and last droplets in a pulse. If the pulse has $N$ droplets, the expectation value of $\Delta t$ is given as \begin{equation} \label{eq:total_nuc_dur}
    \langle\Delta t\rangle = (N-1) \tau_\text{nuc}\ ,
\end{equation} 
since there are N-1 nucleation events from the oldest droplet formed in the pulse until the most recently nucleated droplet, and the average waiting time between these events is $\tau_\text{nuc}$. Therefore, \begin{equation} \label{eq:pulse_width}
    \langle\Delta_\text{pulse}\rangle \simeq v_\text{sed}(N-1)\tau_\text{nuc}
\end{equation}

We speculate that faster nucleation leads to a more synchronous nucleation of droplet at the top, decreasing the pulse dispersion and thus the average pulse width $\langle \Delta_\text{pulse} \rangle$. 
We make nucleation faster by increasing the kinetic nucleation rate coefficient $K$ (Eq.~\eqref{eq:nucleation-rate}). From Eq.~\eqref{eq:pulse_width} and Eq.~\eqref{eq:nucleation-rate} we expect that the average pulse width $\langle \Delta_\text{pulse} \rangle$ decreases as we increase $K$.  In experiments, larger $K$ could be realized through enhancing the adsorption of coacervate components or raft-like heterogeneity of the lipid-aqueous system boundary.  
The variation in the width $\langle\Delta_\text{pulse}\rangle$ with $K$ from simulations shown in Fig~\ref{fig:fig4}(c) shows that $\langle\Delta_\text{pulse}\rangle$ decreases steadily with $K$, pointing towards more synchronized nucleation and sedimentation as expected from the analytical ansatz.

\section{Discussion \& Conclusion}

Identifying the physical principles governing oscillations in active emulsions is challenging because droplet nucleation, growth, sedimentation, fusion, and dissolution are coupled through the concentration fields surrounding the droplets.
Resolving this feedback using conventional phase-field simulations is computationally demanding and often obscures the individual processes controlling the dynamics.
Here, we instead introduced a discrete-droplet framework in which each droplet is characterized by its position and radius. 
In the sharp-interface limit, the concentration profiles inside and outside each droplet determine its instantaneous growth or shrinkage rate, while the shared concentration field mediates interactions between droplets. 
Coupling these dynamics to the low-Reynolds-number flow generated by sedimenting droplets provides a minimal description that retains the essential ingredients of the experimentally observed oscillations~\cite{sastre2025size}.

Our simulations reproduce the complete oscillatory cycle: droplets nucleate, grow and sediment, fuse at the bottom of the system, and subsequently shrink before a new generation of droplets emerges. 
Importantly, the model allows these individual stages to be separated and thereby reveals what sets the timescale of the collective dynamics. 
The oscillation period is approximately the sum of two dominant timescales: the time required for droplets to nucleate at the top system boundary and the time required for the resulting droplets to sediment. 
This simple decomposition quantitatively accounts for oscillation frequencies in the experimentally observed range and captures their dependence on fuel concentration. 
The theory further predicts a non-monotonic dependence of the oscillation frequency on fuel. 
This optimum originates from two competing consequences of chemical activation: increasing fuel promotes production of the strongly phase-separating component, but simultaneously enhances its sequestration into the droplet phase. 
The oscillation frequency therefore reflects a competition between chemical activation, phase partitioning, and transport rather than the intrinsic frequency of a chemical oscillator.

Beyond the oscillation period, the model reveals how individual droplets organize into coherent spatio-temporal pulses. 
The degree of synchronization is controlled by nucleation: slow nucleation distributes droplet birth over an extended time interval and produces a rain of sedimenting droplets, whereas rapid nucleation concentrates droplet formation into a narrow temporal window. 
The resulting population subsequently grows and sediments collectively, producing sharply defined propagating droplet pulses. 
Increasing the nucleation rate thus narrows the pulses and enhances their coherence. 
This identifies nucleation not merely as the process initiating each cycle, but as a control parameter governing the collective synchronization of an active emulsion.

The central result of our work is that sustained oscillations do not require an oscillatory chemical reaction network. 
Instead, stationary chemical driving maintains the emulsion away from phase equilibrium and establishes a preferred droplet size. 
Fusion provides the essential nonlinearity: when sedimenting droplets merge, they are displaced far beyond this stationary size and consequently shrink, releasing material back into the dilute phase. 
Transport of this material generates the supersaturation needed to nucleate the next generation of droplets. 
Together, active size control, fusion, and delayed redistribution of material close a feedback loop and transform an otherwise stationary active emulsion into a gravity-driven oscillator. 
In this sense, the clock is not encoded in the reaction network itself: its period emerges collectively from phase separation and transport.

This distinction suggests a broader route towards temporal organization in chemically driven soft matter. 
Previous synthetic oscillators based on molecular assemblies have often coupled an intrinsically oscillatory chemical reaction to a responsive material, thereby translating chemical oscillations into periodic changes in micelles, vesicles, gels, or other supramolecular assemblies~\cite{Yoshida1996,Ueki2013,LeiraIglesias2018}. 
Here, by contrast, the phase-separated objects participate directly in generating the feedback responsible for oscillations. The required ingredients are comparatively generic. 
Active size control can arise when reactions favor production of droplet material in the dilute phase and its removal inside droplets~\cite{Zwicker2015,Zwicker2017,Weber2019}, while fusion is intrinsic to liquid droplets. 
Directed transport need not necessarily originate from gravity: external fields, flows, or chemotactic and phoretic drift could provide analogous spatial redistribution. 
Even in the absence of directed transport, stochastic encounters and fusion of Brownian droplets could potentially close a similar feedback loop, although the resulting coherence and timescales would differ.

These observations also shift the role of phase separation in synthetic-cell design~\cite{Crowe2018,Martin2019}. 
Phase-separated compartments are often viewed primarily as structures that concentrate and spatially organize molecules and biochemical reactions~\cite{Banani2017,Nakashima2019}. 
Chemically active droplets can instead constitute dynamical elements whose size, number, and spatial organization continuously reorganize in response to material fluxes~\cite{Weber2019,Zwicker2022,Juelicher2024DropletPhysics,Cates2025ActivePhaseSeparation}. 
Our results show that such compartments can additionally generate an intrinsic timescale and synchronize into collective pulses. 
Active emulsions may therefore provide a route for engineering synthetic systems in which spatial compartmentalization and temporal organization arise from the same physical mechanism.

The discrete-droplet framework also provides a route to studying competition and selection in heterogeneous active emulsions. Different droplet populations could be assigned distinct compositions, partition coefficients, reaction rates, and stationary sizes while competing for the same pool of chemical material and fuel. 
Such differences already lead experimentally to composition-dependent growth rates and fitness of active coacervates~\cite{TenaSolsona2021}, while competition for a common chemical feedstock has been shown to select between different self-assembling protocell populations~\cite{Morrow2020}. 
Extending our framework to multiple droplet species could therefore address how reaction-diffusion coupling translates microscopic differences in droplet kinetics into population-level selection. In particular, it could determine whether a faster-growing population necessarily outcompetes others, or whether active size control, material recycling, and spatial transport permit stable coexistence or temporally alternating populations. Such extensions would connect the collective dynamics of active emulsions to emerging questions of competition and selection in synthetic protocell systems~\cite{Lin2023}.

\begin{acknowledgements}
C.A.W. and A.T. thank Hidde Vuijik, Gaetano Granatelli, Filippo Faedi and Thomas Tushar Dutta from Mesoscopic Physics of Life, Univeristy of Augsburg and Francesco Busato from the Boekhoven lab, Technical University of Munich for fruitful discussions, and acknowledge the resources on the LiCCA/ALCC HPC cluster of the University of Augsburg, co-funded by the Deutsche Forschungsgemeinschaft (DFG, German Research Foundation) – Project-ID 499211671. 
Job Boekhoven acknowledges
European Research Council (ERC starting grant 852187, ActiDrops).
C.\ A.\ Weber acknowledges the European Research Council (ERC) under the European Union's Horizon 2020 research and innovation program (fueled Life,  Grant Number 949021).

\end{acknowledgements}

\section*{Declaration of generative AI and AI assisted technologies}
During the preparation of this work, the authors used ChatGPT (OpenAI) for programming assistance, including refining code structure and identifying suitable programming functions, as well as for improving language, readability, and text clarity. Grammarly was also used for grammar, spelling, and language editing. All AI-assisted suggestions were reviewed, verified, and modified where necessary by the authors. The authors take full responsibility for the code, analysis, interpretation, and final content of the manuscript.
\section*{Appendix}
\appendix
\renewcommand{\thesection}{\Alph{section}}

\section{Effective reaction-diffusion kinetics from coupled Cahn-Hilliard/Allen Cahn equations}
\label{apA:Sharp interface limit}
In our model we discussed the kinetics of a three component mixture that can phase separate. The components can also inter-convert into one another by means of a chemical reaction. The spatio-temporal evolution of concentration fields of these components therefore follows coupled Cahn-Hilliard/Allen Cahn equations, the number conserving diffusive kinetics captured by the Cahn-Hilliard part and the non-conserved reaction part by the Allen-Cahn terms, given by \begin{align} \label{eq:Cahn-Hilliard}
    \partial_t c_\alpha = - \nabla\cdot j_\alpha + s_\alpha\ ,
\end{align} where $j_\alpha = -\Lambda \nabla \mu_\alpha$ is the diffusive flux of component $\alpha$ and $s_\alpha$ is the chemical flux \cite{Weber2019}. In case of strong phase separation, when the interfacial width is small compared to droplet radius $R$, we can approximate Eq.~\eqref{eq:Cahn-Hilliard} with a coarse-grained description. We treat the interface to be infinitesimally thin, meaning that it occupies no volume. Since concentration variations in each phase are small, the kinetics within each phase can be approximated by an effective reaction diffusion equation \cite{Weber2019} \begin{equation}
    \partial_t c_\alpha \simeq D_\alpha \nabla^2 c_\alpha + s_\alpha\ ,
 \end{equation}
where \begin{equation}
    D_\alpha = \Lambda(c_\alpha,c_\beta)f''_\alpha\ 
\end{equation} is the effective diffusivity of component $\alpha$, $f$ being the Flory-Huggins free energy density.
 Due to diffusive exchange of molecules between the two phases, the droplet radius $R$ can grow or shrink. The rate of droplet growth is given as \begin{equation}
    \frac{d}{dt}R = \frac{j_\alpha^\text{in}-j_\alpha^\text{out}}{c_\alpha^\text{in}-c_\alpha^\text{out}} \ ,
\end{equation} where $j_\alpha^\text{in/out}$ are diffusive fluxes inside/outside at the droplet interface. 
Therefore, we represent the full spatial kinetics in Eq.~\eqref{eq:Cahn-Hilliard} as effective diffusive kinetics inside and outside droplets. The diffusive exchange between phases is captured by growth or shrinkage of droplets.

\section{Quasi-static approximation and internal concentration profiles of active droplets} 
\label{apB:Quasi-static droplets}
Using the sharp interface limit described in Appendix~\ref{apA:Sharp interface limit}, we can solve for the concentration profiles inside each droplet, allowing us to then track their kinetics by using Eq.~\eqref{eq:Droplet_growth_rate}. For this we need the diffusive fluxes close to the interface $j^\text{in/out}$. To this end, we solve the linearized reaction-diffusion equations inside and outside the droplet. We work in the quasi-static limit, where we assume that concentration profiles relax much faster compared to the interface velocity $\text{d}R/\text{d}t$. We have to solve for the stationary concentration profiles for given droplet radius. We then need two boundary conditions to fully determine the stationary solutions of Eq.~\eqref{eq:quasi-static} in each phase. We denote $r$ to be the radial coordinate around a given droplet, and assume spherically symmetric concentration profiles inside and outside the droplet close to the interface.  Inside the droplet ($0<r<R$), we use no flux condition at the $r=0$ and impose phase equilibrium condition for concentration the interface $c(r=R) = c^\text{in}$. The general form of these solutions inside droplets for $0 < r < R$ is then
\begin{subequations} 
\begin{align}\label{eq:solutions_dense_phase}
      &c_A(r) = \frac{c_A^{\text{in}}\rho_k^{\text{in}}-c_B^{\text{in}} }{\rho_D^{\text{in}} + \rho_k^{\text{in}}} \frac{R\sinh{(r/\lambda^{\text{in}})}}{r\sinh{(R/\lambda^{\text{in}})}} + \frac{c_A^{\text{in}}\rho_D^{\text{in}}+{c_B^{\text{in}}}}{\rho_D^{\text{in}} + \rho_k^{\text{in}}} \ , \\
      &c_B(r) = \rho_D^{\text{in}}\frac{c_B^{\text{in}} - c_A^{\text{in}}\rho_k^{\text{in}}}{\rho_D^{\text{in}} + \rho_k^{\text{in}}} \frac{R\sinh{(r/\lambda^{\text{in}})}}{r\sinh{(R/\lambda^{\text{in}})}} + \rho_k^{\text{in}} \frac{c_A^{\text{in}}\rho_D^{\text{in}}+{c_B^{\text{in}}}}{\rho_D^{\text{in}} + \rho_k^{\text{in}}} \ ,
    \end{align} 
\end{subequations}
\cite{Bergmann2023}
where $\rho_D^\text{in} = D_A^\text{in}/D_B^\text{in}$ and $\rho_k^\text{in} = k_\text{act}c_F^\text{in}/k_\text{deact}$, $\lambda^\text{in}$ is the reaction diffusion length scale given by \begin{equation} \label{RD-lengthscale_dense}
    \lambda^\text{in} = \sqrt{\frac{D_A^\text{in}D_B^\text{in}}{D_B^\text{in}k_\text{act}c_F^\text{in}+D_A^\text{in}k_\text{deact}}}\ .
\end{equation} and $c_A^\text{in}$ and $c_B^{in}$. Similarly, outside the droplet ($R<r<R_\text{ref}$), where $R_\text{ref}$ is a point outside the droplet close to the interface. We have phase equilibrium for $c(r=R) = c^\text{out}$, and Dirichlet boundary condition $c(r = R_\text{ref}) = c^\text{ref}$. Assuming $D_A^\text{out} = D_B^\text{out}$, the solutions then take the form  
\begin{align}
\label{eq:solutions_dil_phase}
         c_A(r) &=  \nonumber  \frac{c_A^\text{ref}k_\text{act}c_F^\text{out}-c_B^\text{ref}k_\text{deact}}{k_\text{act}c_F^\text{out}+k_\text{deact}\rho_D^{\text{out}}} \frac{R_\text{ref}}{r}\frac{\sinh{[(R-r)/\lambda^{\text{out}}]}}{\sinh{[(R-R_\text{ref})/\lambda^{\text{out}}]}}  \\ \nonumber & + \frac{c_A^\text{out}k_\text{act}c_F^\text{out}-c_B^\text{out}k_\text{deact}}{k_\text{act}c_F^\text{out}+k_\text{deact}} \frac{R}{r}\frac{\sinh{[(R_\text{ref}-r)/\lambda^{\text{out}}]}}{\sinh{[(R_\text{ref}-R)/\lambda^{\text{out}}]}}  \\ \nonumber &  +\frac{(c_A^\text{out}+c_B^\text{out})}{1+\rho_k^{\text{out}}} \\  &+ \frac{(R/r-1)(c_A^\text{out}+c_B^\text{out}-c_A^\text{ref}-c_B^\text{ref})}{(1+\rho_k^{\text{out}})(1-R/R_\text{ref})}\, , 
    \end{align}
 \begin{align}
        c_B(r) &=  \nonumber  -\frac{c_A^\text{ref}k_\text{act}c_F^\text{out}-c_B^\text{ref}k_\text{deact}}{k_\text{act}c_F^\text{out}+k_\text{deact}\rho_D^{\text{out}}} \frac{R_\text{ref}}{r}\frac{\sinh{[(R-r)/\lambda^{\text{out}}]}}{\sinh{[(R-R_\text{ref})/\lambda^{\text{out}}]}}  \\ \nonumber & - \frac{c_A^\text{out}k_\text{act}c_F^\text{out}-c_B^\text{out}k_\text{deact}}{k_\text{act}c_F^\text{out}+k_\text{deact}} \frac{R}{r}\frac{\sinh{[(R_\text{ref}-r)/\lambda^{\text{out}}]}}{\sinh{[(R_\text{ref}-R)/\lambda^{\text{out}}]}} \\ \nonumber &+ \frac{\rho_k^{\text{out}}(c_A^\text{out}+c_B^\text{out})}{1+\rho_k^{\text{out}}} \\ & + \frac{\rho_k^{\text{out}}(R/r-1)(c_A^\text{out}+c_B^\text{out}-c_A^\text{ref}-c_B^\text{ref})}{(1+\rho_k^{\text{out}})(1-R/R_\text{ref})}\ ,
    \end{align} 
   
$\rho_k^\text{out} = k_\text{act}c_F^\text{out}/k_\text{deact}$, $\lambda^\text{out}$ is the reaction diffusion length scale given by 
\begin{equation} \label{RD_lengthscale_dil}
    \lambda^\text{out} = \sqrt{\frac{D^\text{out}}{k_\text{act}c_F^\text{out}+k_\text{deact}}}\ , \end{equation}

The phase equilibrium condition at the interface ensures 
\begin{subequations} \label{eq:equil_conditions}
    \begin{align} \mu_A^\text{in}(c_A^\text{in},c_B^\text{in}) &= \mu_A^\text{out}(c_A^\text{out},c_B^\text{out}) \\ \mu_B^\text{in}(c_A^\text{in},c_B^\text{in}) &= \mu_B^\text{out}(c_A^\text{out},c_B^\text{out}) \\ \Pi^\text{in}(c_A^\text{in},c_B^\text{in}) &= \Pi^\text{out}(c_A^\text{out},c_B^\text{out}) + \frac{2\gamma}{R} \ .\end{align}
\end{subequations}

In addition, the droplet radius in Eq.~\eqref{eq:Droplet_growth_rate} must be identical for both components, or \begin{equation}\label{eq:interface_velocity_constraint}
    \frac{j_A^\text{in}-j_A^\text{out}}{c_A^\text{in}-c_A^\text{out}} = \frac{j_B^\text{in}-j_B^\text{out}}{c_B^\text{in}-c_B^\text{out}}\ .
\end{equation}

 The fluxes $j^\text{out} = -D^\text{out}\partial_r c $ directly depend on  $c_A^\text{ref}$ and $c_B^\text{ref}$ through Eq.~\eqref{eq:solutions_dil_phase}. Therefore the growth rate of droplets $dR/dt$ from Eq.~\eqref{eq:Droplet_growth_rate} is governed by the local composition of the background field $c_A^\text{ref}$ and $c_B^\text{ref}$. 
Typically droplets grow in a region rich in $c_A$ and $c_B$ and shrink when $A$ and $B$ are depleted. Each droplet senses its immediate environment through $c_A^\text{ref}$ and $c_B^\text{ref}$ and is coupled to other droplets in the system through these fields. 

As discussed previously, for simplicity of modelling, we consider complete wetting of droplets at the bottom surface, creating a wetted layer of height $h$, which can grow or shrink as per Eq.~\eqref{eq:height-growth-rate}. The concentration profiles inside and outside close to the interface of this layer have been derived in the quasi-static limit in the Appendix~\ref{apC:Quasi-static layer}. 

Given that the system is active, this can lead to interesting collective behavior such as oscillations or size control, depending on the experimental conditions.

\section{Quasi-static approximation and internal concentration profiles for wetted layer}
\label{apC:Quasi-static layer}
Just like in Appendix~\ref{apB:Quasi-static droplets} where we solve for quasi-static concentration profiles for droplets, here we present the transient quasi-static solutions for the wetted layer. Since the layer height is assumed to be uniform in the $X$ and $Y$ directions, the only degree of freedom is $Z$, effectively making it a one dimensional reaction-diffusion problem. In other words, we solve for $0 \simeq D_\alpha^\text{in/out} \partial_z^2 c_\alpha(z)\  + r(\{c_\beta(z)\})$ inside and outside the layer interface. 

Like the droplets, we have the boundary conditions Eq.~\eqref{eq:equil_conditions} ensuring phase equilibrium at the layer interface, with the exception that for a flat layer, the curvature $R^{-1}\to 0$, so the Laplace pressure $2\gamma/R$ drops out and $\Pi^\text{in} = \Pi^\text{out}$. Since we have a closed box, $\partial_zc_\alpha(z=0) = 0$. To determine the kinetics of the layer height, we choose a reference point $z_\text{ref}$ outside the layer, with concentrations $c_\alpha^\text{ref}$, which provides the second boundary condition for concentration profiles outside the layer interface, $c_\alpha(z=z_\text{ref}) = c_\alpha^\text{ref}$. The quasi-static solutions for a layer of height $h$ in $0<z<h$ are \begin{align}
    c_A(z) &= \frac{k_\text{deact}c_B^\text{in} - k_\text{act}c_F^\text{in}c_A^\text{in}}{k_\text{act}c_F^\text{in}+  k_\text{deact}\rho_D^\text{in}} \left[1-\frac{\cosh(z/\lambda^\text{in})}{\cosh(h/\lambda^\text{in})}\right] + c_A^\text{in} \\ \nonumber c_B(z) &= \frac{\rho_D^\text{in}(k_\text{deact}c_B^\text{in} - k_\text{act}c_F^\text{in}c_A^\text{in})}{k_\text{act}c_F^\text{in}+  k_\text{deact}\rho_D^\text{in}} \left[\frac{\rho_k^\text{in}}{\rho_D^\text{in}}+\frac{\cosh(z/\lambda^\text{in})}{\cosh(h/\lambda^\text{in})}\right]  \\ \nonumber &  + c_A^\text{in}\rho_k^\text{in}\ ,
\end{align} and the solutions in $h<z<z_\text{ref}$ are \begin{align}
    c_A(z) &= \frac{k_\text{deg}}{D_A^\text{out}} \cdot \frac{1}{\sinh((L-h)/\lambda^\text{out})} \cdot \\ \nonumber &\cdot \left(-C_1^\text{ref} \sinh((z-h)/\lambda^\text{out}) - C_1^\text{out}\sinh((L-z)/\lambda^\text{out})\right)\\ \nonumber
    & + \frac{k_\text{deg}}{D_B^\text{out}}\cdot \left[\frac{z-h}{L-h}\cdot (C_2^\text{ref}-C_2^\text{out}) + C_2^\text{out}\right]\ , \\ \nonumber 
    c_B(z) &= -\frac{k_\text{deg}}{D_B^\text{out}} \cdot \frac{1}{\sinh((L-h)/\lambda^\text{out})} \cdot \\ \nonumber &\cdot \left(-C_1^\text{ref} \sinh((z-h)/\lambda^\text{out}) - C_1^\text{out}\sinh((L-z)/\lambda^\text{out})\right) \\ \nonumber
    & + \frac{k_\text{act}\cdot c_F^\text{out}}{D_B^\text{out}}\cdot \left[\frac{z-h}{L-h}\cdot (C_2^\text{ref}-C_2^\text{out}) + C_2^\text{out}\right]\ ,
    \end{align}
where \begin{align*} 
    C_1^\text{ref} &= (\lambda^\text{out})^2 (c_B^\text{ref}-\rho_k^\text{out}c_A^\text{ref})\\ 
    C_1^\text{out} &= (\lambda^\text{out})^2 (c_B^\text{out}-\rho_k^\text{out}c_A^\text{out})\\ 
    C_2^\text{ref} &= (\lambda^\text{out})^2 (c_B^\text{ref}/\rho_D^\text{out}+c_A^\text{ref})\\ 
   C_2^\text{out} &= (\lambda^\text{out})^2 (c_B^\text{out}/\rho_D^\text{out}+c_A^\text{out}) \ ,
\end{align*}

and $\rho_D$, $\rho_k$ and $\lambda$ are as defined before. Similarly, we can determine the concentration profiles when the system is in a stationary state, when the layer height does not change in time. For such a situation, the solutions inside the layer remain unchanged, since the boundary conditions remain the same as before. However, outside the layer, we can use Neumann boundary condition at the system boundary, since the system is a closed box. This gives us the following solutions for a stationary layer at any point outside the layer $h_\text{stat}<z<H$, where $H$ is the system height \begin{align}
    c_A(z) &= (\lambda^\text{out})^2 \frac{k_\text{deg}}{D_A^\text{out}} (\rho_k^\text{out}c_A^\text{out}-c_B^\text{out}) \frac{\cosh ((H-z)/\lambda^\text{out})}{\cosh ((H-h)/\lambda^\text{out})} \\ \nonumber &+ \frac{1}{\rho_k}\left(c_B^\text{out}-(\lambda^\text{out})^2 \frac{k_\text{deg}}{D_B^\text{out}}(c_B^\text{out}-\rho_k^\text{out}c_A^\text{out})\right) \\ \nonumber
    c_B(z) &= (\lambda^\text{out})^2 \frac{k_\text{deg}}{D_B^\text{out}} (c_B^\text{out}-\rho_k^\text{out}c_A^\text{out}) \frac{\cosh ((H-z)/\lambda^\text{out})}{\cosh ((H-h)/\lambda^\text{out})} \\ \nonumber &+ \left(c_B^\text{out}-(\lambda^\text{out})^2 \frac{k_\text{deg}}{D_B^\text{out}}(c_B^\text{out}-\rho_k^\text{out}c_A^\text{out})\right)\ .
\end{align}

In addition to this, the concentration profiles for a stationary layer depend on the total amount of material in the system $\psi$, which enters as a boundary condition when solving for the unknowns $c_\alpha^\text{in/out}$ and $h_\text{stat}$ along with Eq.~\eqref{eq:equil_conditions}.

\section{Amplitude of capillary fluctuations for droplets}
\label{apD:Capillary fluctuations}
Amplitude of capillary fluctuations for a spherical droplet is of the order $\sim \sqrt{k_BT/\gamma}$ \cite{milner1987dynamical}, where $\gamma$ is the droplet's surface tension. We consider a surface tension of $100 \mu Nm^{-1}$ for our droplets, which is in the range observed for coacervate droplets \cite{spruijt2009interfacial}. For this surface tension, the capillary fluctuation amplitude at room temperature turns out to be of the order of $nm$. We can therefore safely neglect capillary fluctuations. 

\section{Estimates of Reynolds number, P\'eclet number, Weber number, Capillary number and Bond number for droplets}
\label{apE: dimensionless numbers}

\subsubsection*{Reynolds number} The Reynolds number is the ratio of inertial to viscous forces, $Re = \rho^\text{out} v_\text{sed}R/\eta$~\cite{batchelor2000introduction}, associated with a droplet of radius $R$ sedimenting at speed $v_\text{sed}$.  For very low Reynolds number ($Re\ll 1 $), the inertial terms in droplet's center of mass motion and also in the flow field dynamics can be neglected. In our system, the fluid density outside droplets is $\rho^\text{out} \approx 1g\ cm^{-3} $, the droplet sedimentation velocity $v_\text{sed} \approx 1 \mu ms^{-1}$, droplet size $R \approx 1 \mu m$ and fluid viscosity outside droplets $\eta \approx 5\ mPa\cdot{}s$. This gives a Reynolds number of $10^{-6}$, which means we are well-posed to assume overdamped dynamics. 
\subsubsection*{P\'eclet number} The P\'eclet number $Pe = v_\text{sed} R_i/D_\text{drop}$~\cite{stewart1962transport} quantifies the propensity of sedimentation compared to Brownian diffusion, for a single droplet of radius $R_i$. As before, we consider the droplet sedimentation velocity $v_\text{sed} \approx 1 \mu ms^{-1}$. The droplet diffusivity is given by $D_\text{drop} = k_BT / 6\pi \eta R_i$ obeying the Sutherland-Einstein-Stokes relationship. Nucleated droplets are usually of the order of a few nanometers. For $R_i \sim 10 nm$, the P\'eclet number turns out to be of the order $10^{-7}$, meaning that for newly nucleated droplets sedimentation is irrelevant. The P\'eclet number $Pe \approx 1$ for $R_i \approx 0.6 \mu m$, which means droplets bigger than $0.6 \mu m$ are sedimentation dominated. 

\subsubsection*{Weber number, Capillary number and Bond number} While sedimenting, droplets are subject to hydrodynamic deforming forces due to the relative motion of fluid surrounding them. These forces include inertial forces, viscous forces and gravity. The surface tension $\gamma$ of droplets fights against these stresses, therefore it is meaningful to look at the ratio of these forces relative to the restoring force provided by the surface tension. The dimensionless number representing the ratio of inertial force to surface tension is the Weber number $\textit{We} = 2\rho^\text{out}v_\text{sed}^2 R/\gamma$~\cite{clift2005bubbles}. The surface tension of our droplets was considered to be around $100 \mu Nm^{-1}$, which is in the range observed for coacervate droplets~\cite{spruijt2009interfacial}. For typical droplet sizes and sedimentation speeds in our system, we get that $\textit{We}\sim 10^{-11}$, meaning that droplets remain almost perfectly spherical as they sediment. Similarly, for viscous force we have the Capillary number $Ca = \eta v_\text{sed}/\gamma$. For our system, $Ca \sim 10^{-5}$, meaning that viscous deformation of droplets can also be neglected.  Lastly, hydrostatic pressure difference across the droplet interface varies with the $z-$coordinate because of the density difference between droplet interior and exterior. This can make the droplet curvature $z-$dependent. This tendency is captured by the Bond number $Bo = (\rho^\text{in}-\rho^\text{out})gR^2/\gamma$. For the parameters mentioned above, $Bo \sim 10^{-5}$, so hydrostatic deformation of droplets is negligible.

\section{Derivation of nucleation rate for a ternary mixture}
\label{apF:Nuc_rate}
We know from Kramer theory and from reference \cite{hanggi1990reaction}, the nucleation time of droplets is governed by the height of the free energy barrier $\Delta \tilde{F}$ associated with formation of a droplet of critical radius $R_c$, and is given by \begin{equation}
    \tau_\text{nuc} = \tau_0 \exp(\Delta \tilde{F}/k_BT)\ ,
\end{equation} where $\tau_0$ is a kinetic coefficient. We now derive an expression for $\Delta \tilde{F}$ for a ternary mixture. We expand the free energy around phase equilibrium to get $\Delta \tilde{F}$, as derived for a binary mixture in \cite{ziethen2023nucleation}. We start with a homogeneous mixture of composition $(c_A^\text{avg},c_B^\text{avg})$. When a droplet of volume $V_\text{drop}$ is formed, the total free energy of the system of volume $V_\text{sys}$ can be given as \begin{equation}
    F = V_\text{drop}\cdot f(c_A^\text{in},c_B^\text{in}) + (V_\text{sys}-V_\text{drop})\cdot f(c_A^\text{out},c_B^\text{out}) + \gamma A\ ,
\end{equation}
where $f$ is the Flory-Huggins free energy density, $\gamma$ is the surface tension and $A$ is the droplet surface area. Now assuming that the average composition of the mixture is close to the dilute binodal, we can write upto first order \begin{align}
    f(c_A^\text{out},c_B^\text{out}) &= f(c_A^\text{avg},c_B^\text{avg}) + \\  \nonumber &\mu_A^\text{avg}\cdot(c_A^\text{out}-c_A^\text{avg}) + \mu_B^\text{avg}\cdot(c_B^\text{out}-c_B^\text{avg})\ ,
\end{align} 
where $\mu_\alpha^\text{avg} = \partial f/\partial c_\alpha |_{(c_A^\text{avg},c_B^\text{avg})}$ is the exchange chemical potential of component $\alpha$ at the average composition. The total free energy then becomes \begin{align}
    F &= V_\text{drop}\cdot f(c_A^\text{in},c_B^\text{in}) + (V_\text{sys}-V_\text{drop})\cdot \\ \nonumber & \left(f(c_A^\text{avg},c_B^\text{avg}) +  \mu_A^\text{avg}\cdot(c_A^\text{out}-c_A^\text{avg}) +  \mu_B^\text{avg}\cdot(c_B^\text{out}-c_B^\text{avg})\right)  \\  \nonumber& +\gamma A\ .
\end{align}
Using lever rule, we can express \begin{equation} c_\alpha^\text{out} = \frac{V_\text{sys}c_\alpha^\text{avg}-V_\text{drop}c_\alpha^\text{in}}{V_\text{sys}-V_\text{drop}}\ .
\end{equation}
The total free energy then simplifies to \begin{align} \label{eq:tot-free_en-nuc}
    F &= V_\text{sys} f(c_A^\text{avg},c_B^\text{avg}) - \Delta f_\text{bulk}V_\text{drop}  \\ \nonumber & + \gamma A\ ,
\end{align}

where \begin{align} \label{eq:bulk_free_en}
     \Delta f_\text{bulk} &= f(c_A^\text{avg},c_B^\text{avg})-f(c_A^\text{in},c_B^\text{in}) \\ \nonumber & + \mu_A^\text{avg}(c_A^\text{in}-c_A^\text{avg}) + \mu_B^\text{avg}(c_B^\text{in}-c_B^\text{avg})
\end{align} is the bulk free energy gain term. 

$V_\text{sys} f(c_A^\text{avg},c_B^\text{avg})$ is the initial free energy before droplet nucleation. Subtracting this from the total free energy in Eq.~\eqref{eq:tot-free_en-nuc}, the free energy change associated with droplet nucleation is given as \begin{equation} \label{eq:deta_F_nuc}
    \Delta F = -\Delta f_\text{bulk} V_\text{drop} + \gamma A\ .
\end{equation}  

In terms of droplet radius, 
\begin{equation} 
    \Delta F = -\Delta f_\text{bulk} 4\pi R^3/3 + \gamma 4\pi R^2\ .
\end{equation}  From the above equation, it is clear to see that for small radii $R$, the quadratic term dominates, meaning that the surface free energetic cost to nucleate a droplet is higher than the bulk free energy gain. However, after a certain radius $R_c$, the cubic term takes over and $\Delta F$ decreases for $R>R_c$. 

The barrier height $\Delta\tilde{F}$ and critical radius $R_c$ are defined by \begin{equation}
    \partial \Delta F/\partial R = 0\ ,
\end{equation}
giving \begin{align}R_c &= 2\gamma /\Delta f_\text{bulk} \, , \\ \Delta\tilde{F} &= 16\pi \gamma^3/3\Delta f_\text{bulk}^2 \ .\end{align} Droplets with $R<R_c$ shrink in size since $\partial \Delta F/\partial R > 0$ for $R<R_c$, and droplets above the critical radius grow. 
Now, let us simplify the bulk free energy gain per unit volume from Eq.~\eqref{eq:bulk_free_en}. We can expand $f(c_A^\text{avg},c_B^\text{avg})$ as \begin{align} 
    f(c_A^\text{avg},c_B^\text{avg}) &= f(c_A^\text{out},c_B^\text{out} )\\ \nonumber & + \mu_A^\text{avg}(c_A^\text{avg}-c_A^\text{out}) + \mu_B^\text{avg}(c_B^\text{avg}-c_B^\text{out})\ .
\end{align}

Substituting that in Eq.~\eqref{eq:bulk_free_en}, we get \begin{align}
    \Delta f_\text{bulk} &= f(c_A^\text{out},c_B^\text{out} ) - f(c_A^\text{in},c_B^\text{in} ) \\ \nonumber &+ \mu_A^\text{avg}(c_A^\text{in}-c_A^\text{out}) + \mu_B^\text{avg}(c_B^\text{in}-c_B^\text{out})\ .
\end{align}

From the common tangent rule, we can write \begin{align} 
    f(c_A^\text{in},c_B^\text{in}) &= f(c_A^\text{out},c_B^\text{out} ) \\ \nonumber &+ \mu_A^\text{eq}(c_A^\text{in}-c_A^\text{out}) + \mu_B^\text{eq}(c_B^\text{in}-c_B^\text{out})\ ,
\end{align}
where $\mu_\alpha^\text{eq}$ are chemical potentials at phase equilibrium.

This gives us \begin{align}
    \Delta f_\text{bulk} &= \mu_A^\text{avg}(c_A^\text{in}-c_A^\text{out}) + \mu_B^\text{avg}(c_B^\text{in}-c_B^\text{out})\ \\ \nonumber \quad &- \mu_A^\text{eq}(c_A^\text{in}-c_A^\text{out}) - \mu_B^\text{eq}(c_B^\text{in}-c_B^\text{out})
    \\  \nonumber &
    = (c_A^\text{in}-c_A^\text{out})(\mu_A^\text{avg}-\mu_A^\text{eq})   \\ \nonumber & \quad +( c_B^\text{in}-c_B^\text{out})(\mu_B^\text{avg}-\mu_B^\text{eq})\ .
\end{align} 

Since the average composition of the mixture is close to the dilute binodal, we can express to the first order \begin{align}
    \mu_\alpha^\text{avg} &= \mu_\alpha^{eq} + \sum_\beta\partial \mu_\alpha/\partial c_\beta |_{c_\beta = c_\beta^\text{out}}\cdot(c_\beta^\text{avg}-c_\beta^\text{out})\ , \\ \nonumber  \mu_\alpha^\text{avg} - \mu_\alpha^\text{eq} &=  \sum_\beta\partial^2f/\partial c_\alpha c_\beta \cdot(c_\beta^\text{avg}-c_\beta^\text{out})\ .
\end{align}

That gives us \begin{align}
    \Delta f_\text{bulk} &= \sum_\alpha(c_\alpha^\text{in}-c_\alpha^\text{out}) \sum_\beta\partial^2f/\partial c_\alpha c_\beta \cdot(c_\beta^\text{avg}-c_\beta^\text{out})\ . 
\end{align}

Defining $\epsilon_\alpha = (c_\alpha^\text{avg}-c_\alpha^\text{out})/c_\alpha^\text{out}$, we can write \begin{equation*}
    \Delta f_\text{bulk} = \sum_\alpha \sum_\beta (c_\alpha^\text{in}-c_\alpha^\text{out}) f''_{\alpha\beta} \epsilon_\beta c_\beta^\text{out}
\end{equation*}

We can then define an effective supersaturation for the ternary mixture in terms of $\epsilon_\alpha$ as \begin{equation} \label{eq:supersat}
    \epsilon = \sum_\alpha \sum_\beta  \frac{f''_{\alpha\beta}c_\beta^\text{out}\epsilon_\beta (c_\alpha^\text{in}-c_\alpha^\text{out})}{c_B^\text{out}(c_B^\text{in}-c_B^\text{out})f''_{B}} \ ,
\end{equation}

so that the expression for bulk free energy gain becomes \begin{equation}
    \Delta f_\text{bulk} = c_B^\text{out}(c_B^\text{in}-c_B^\text{out})f''_{B} \cdot \epsilon \ .
\end{equation}

And therefore, the nucleation time $\tau_\text{nuc}$ takes the familiar form from Eq.~\eqref{eq:nucleation-rate}\begin{equation} \label{eq:2D_tau_nuc}
    \tau_\text{nuc} = \tau_0 \exp\left[16\pi \gamma^3/\left(3(c_B^\text{out})^2(c_B^\text{in}-c_B^\text{out})^2(f''_{B})^2 \epsilon^2 k_BT\right)\right]\ ,
\end{equation} 

where $\tau_0 = K^{-1}$, and $K$ is the nucleation rate constant from Eq.~\eqref{eq:nucleation-rate}
We note that the nucleation rate depends on the composition of the mixture as a whole, including both $c_A$ and $c_B$. 

\section{Summary of model assumptions}
\label{apG:model summary}
We use the following set of assumptions for which our  theoretical description holds: 

\begin{itemize}
    \item Droplet number density are in the range of 1/(100 $\mu m^3$) with typical droplet size on the micron scale, so we assume that droplets are  sparsely distributed.
    \item Droplets remain spherical during sedimentation (discussions see Appendix~\ref{apE: dimensionless numbers}).
    \item Droplets experience a drag force given by the Stokes equation $F = 6\pi \eta R v_\text{sed}$, similar to a colloidal sphere of radius $R$ (see Appendix~\ref{apE: dimensionless numbers}).
    \item The system is always in the low Reynolds number regime, so that flows outside droplets are described the Stokes equation~\eqref{eq:Stokes} (Appendix~\ref{apE: dimensionless numbers}).
    \item Droplet interface width is much thinner than droplet radius $R_i$. 
    \item Droplet interface moves much slower compared to the relaxation of concentration profiles, allowing us to work in the quasi-static limit $\partial_t c_\alpha \approx 0$.
    \item Droplets on contact with the bottom surface of the micro-reactor completely wet the surface to form a wetted layer of dynamic height $h(t)$.  
    
\end{itemize}

\section{Numerical integration and parameter table}\label{appH:num_int_param}

In order to track the evolution of the system in time, we use the following numerical scheme. We consider a cuboidal micro-reactor of size $L \times L \times H$. We use a coarse cubical grid of dimension $d\text{x}$, and define all concentration fields and velocity field of the solvent-rich phase on the grid. For the concentration fields $c_\alpha$ we then solve the reaction-diffusion equation Eq.~\eqref{eq:reaction-diffusion} with Neumann boundary condition for components $A,B$ and Dirichlet boundary condition on all faces of the micro-reactor for the fuel $F$, using finite difference. 
For the velocity field $\textbf{v}(\textbf{x},t)$, we find numerical solutions to Stokes equation Eq.~\eqref{eq:Stokes} by using iterative relaxation at every time step. We use a projection method to impose incompressibility of the flow. We impose no slip boundary condition and treat the micro-reactor walls as hard walls. 

For droplet nucleation, we use a grid-based kinetic Monte Carlo. Each grid block has an associated concentration $c_A$ and $c_B$, which determines the supersaturation $\epsilon$ from Eq.~\eqref{eq:epsilon-ternary}. We can calculate the nucleation rate $\tau_\text{nuc}^{-1}$ using Eq.~\eqref{eq:nucleation-rate} for each grid point. In every time step fixed $dt$, each grid block in the top layer is assigned a random number from $0$ to $1$. If this random number $RN < \tau_\text{nuc}^{-1}\cdot dt$, every grid block that satisfies this condition nucleates a droplet with radius $R_c = 2\gamma/g + \delta$, where $\gamma$ is the surface tension, $g$ is the free energy gain per unit volume, $\delta$ is an arbitrary small number.

The droplets upon nucleation continue to grow as per Eq.~\eqref{eq:Droplet_growth_rate}. The droplets deplete material from the solvent-rich phase. If the droplet growth rate at every time step is $dR/dt$, the rate of change of droplet volume is $4\pi R^2dR/dt$. If the concentrations at the droplet interface inside/outside are $c_\alpha^\text{in/out}$, then at each time step $dt$, the concentration of component $\alpha$ is depleted by $dc_\alpha = 4\pi R^2(dR/dt)\cdot (c_\alpha^\text{in}-c_\alpha^\text{out})\cdot dt/d\text{x}^3$ from the grid block nearest to the droplet. The material then gets redistributed in the subsequent time step by diffusive kinetics. Similar to the droplets, the wetted layer also evolves as per Eq.~\eqref{eq:height-growth-rate}, depleting from or releasing material into the solvent-rich phase. 

The fluxes $j_\alpha^\text{in/out}$ are determined by solving Eq.~\eqref{eq:quasi-static} inside droplets with the boundary conditions Eq.~\eqref{eq:equil_conditions} at the droplet interface. Since we have a system of non-linear coupled equations for the boundary conditions, we use a numerical solver to determine the solutions to these conditions. We have four unknowns $c_\alpha^\text{in/out}$ are four equations, including Eq.~\eqref{eq:equil_conditions} together with the interface velocity condition \begin{equation} \label{eq:interface BC}
    \frac{j_A^\text{in}-j_A^\text{out}}{c_A^\text{in}-c_A^\text{out}} = \frac{j_b^\text{in}-j_B^\text{out}}{c_B^\text{in}-c_B^\text{out}}\ .
\end{equation} 

Together, these allow us to self-consistently evolve the full system with multiple sedimenting droplets in the sharp interface limit, coupled to the solvent-rich phase. 

To track the width of droplet pulses, we decided to not consider the first moment of the droplet number probability density $p(z,t)$; instead, we calculate pulse width $\Delta_\text{pulse}$ as the support of the number distribution around the distribution's maxima decaying to $1/e$-$th$ probability of this maximum. To track $\Delta_\text{pulse}$, we use a python function to identify peaks $p^\text{max}$ of pulses, around which $p(z,t)$ is distributed. We define $\Delta_\text{pulse}$ as the distance between points where $p(z,t)$ falls to $p^\text{max}/e$ on either sides of the maximum. To reduce statistical noise in $p(z,t)$, we track each pulse as it moves through the central region of the micro-reactor from $z = 3H/4$ to $z = H/4$, where $H$ is the micro-reactor height. We determine an average distribution $p^\text{avg}(z)$ by superimposing $p(z,t)$ for different times $t$ as the pulse moves through the central region. We ensure that the average is taken by overlaying the peak $p^\text{max}$ of the different $p(z,t)$. 
We use the same process for each droplet pulse that forms and sediments over the course of the simulation. We then average over the widths of multiple pulses to give us a mean and error associated with $\Delta_\text{pulse}$ for that set of parameters.  
The following table lists the values of all parameters used in numerical simulations of the system.

\begin{table*}[h]
\begin{center}
    
    \begin{tabular}{|c|c|c|}
    \hline
     Quantity  & Symbol & Value \\ \hline
     Interaction parameter precursor-product & $\chi_{AB}$ & -0.18 \\ 
     Interaction parameter precursor-solvent & $\chi_{AS}$ & 0.77 \\ 
     Interaction parameter product-solvent & $\chi_{BS}$ & 0.81 \\ 
     Precursor relative molecular volume & $r_{A}$ & 30 \\ 
     Product relative molecular volume & $r_{B}$ & 60 \\ 
     Surface tension & $\gamma$ & 100 $\mu Nm^{-1}$ \\
     \hline
     Activation rate  & $k_\text{act}$ & $0.17 M^{-1}s^{-1}$ \\
     Fuel partition coefficient inside droplets & $P=c_F^\text{in}/c_F^\text{out}$ & $7$\\
     Deactivation rate  & $k_\text{deact}$ & $0.012s^{-1}$ \\
     Diffusion coefficient of A outside the drop & $D_A^{\text{out}}$ & $300 \mu m^2s^{-1}$ \\
     Diffusion coefficient of B outside the drop & $D_B^{\text{out}}$ & $300 \mu m^2s^{-1}$ \\
     Diffusion coefficient of A inside the drop & $D_A^{\text{in}}$ & $0.04 \mu m^2s^{-1}$ \\
     Diffusion coefficient of B inside the drop & $D_B^{\text{in}}$ & $0.0073 \mu m^2s^{-1}$ \\
     Viscosity of solvent-rich phase & $\eta$ & $5\ mPa\cdot s$ \\
     Mass density of solvent-rich phase & $\rho^\text{out}$ & $1 g\cdot cm^{-3}$ \\
     Mass density of droplets & $\rho^\text{in}$ & $1.8 g\cdot cm^{-3}$ \\
     \hline
     \end{tabular}
\caption{\textbf{ Table with input parameters} used in numerical calculations of the sharp interface model.}
\label{Tab:Eff_drop_model}
\end{center}
\end{table*}

\bibliography{bib}

\end{document}